\documentclass[twocolumn]{aa}
\usepackage{graphicx,psfig}

\newcommand{\kms}{$\mathrm{\,km\,s}^{-1}$ }
\begin{document}

\title{Abundance trends in kinematical groups of the Milky Way's disk.
\thanks{Data available in electronic form at the CDS (Strasbourg, France)}} 
 \author{C. Soubiran 
 \inst{1}
 \& P. Girard
 \inst{1}                     
\authorrunning{Soubiran \& Girard}
          }
\offprints{Caroline Soubiran, soubiran@obs.u-bordeaux1.fr}
\institute{Observatoire de Bordeaux, BP 89, 33270 Floirac, France}

\titlerunning{Abundances trends in the disk}

   \date{\today / accepted}
   
\abstract{
We have compiled a large catalogue of metallicities and abundance ratios 
from the literature in
order to investigate abundance trends of several alpha and iron peak elements
in the thin disk and the thick disk of the Galaxy. The catalogue 
includes 743 stars with abundances of Fe, O, Mg, Ca, Ti, Si, Na, Ni and Al
in the metallicity range -1.30 $<$ [Fe/H] $<$ +0.50. 
We have checked that systematic differences between abundances measured
in the different studies were lower than random errors before combining
them.
Accurate distances and proper motions from Hipparcos, and radial velocities 
from several sources have been retreived for 639 stars and their
velocities (U,V,W) and galactic orbits have been computed. Ages of 322 stars 
have been estimated with the Bayesian method of
isochrone fitting developped by Pont \& Eyer (2004).
Two samples kinematically representative of 
the thin and thick disks have been selected, taking into account
the Hercules stream which is intermediate in kinematics, but with a probable dynamical
origin. 
Our results show that the two disks are chemically well separated, they
overlap greatly in metallicity and both show parallel decreasing trends
of alpha elements with increasing metallicity, in the interval -0.80 $<$ 
[Fe/H] $<$ -0.30.
The Mg enhancement with respect to Fe of the thick disk is measured to be 0.14 dex. An even 
larger enhancement is observed for Al. The thick disk is clearly older than 
the thin disk with a tentative evidence of an AMR over 2-3 Gyr and a hiatus
in star formation before the formation of the thin disk. We do not observe 
a vertical gradient in the metallicity of the thick disk.
The Hercules stream have properties similar to that
of the thin disk, with a wider range of metallicity. Metal-rich stars assigned 
to the thick disk and super metal rich stars assigned to the thin disk appear
as outliers in all their properties.
   }
\maketitle 
  
\section{Introduction}
It is now well admitted that the stellar disk of our Galaxy is made of at least 
two components : the thin disk and the thick disk. The properties
and origin of the thick disk are still matter of debate and require more
investigation. In
the solar neighbourhood, the observed distribution of stellar velocities 
is well approximated by the sum of two distinct velocity ellipsoids 
suggesting that the thin disk and the thick disk are discrete populations.
In the last decade a number of studies have demonstrated that
they also have distinct chemical distributions
(Gratton et al. 1996, Fuhrmann 1998, Bensby et al. 2003,
Mishenina et al. 2004 for instance). Unfortunatly, on the one hand, the 
statistics involving abundance 
ratios in the thick disk are rather poor, on the other hand, all these studies
have not used the same criteria to define the thin disk and the thick disk
populations. This paper is devoted to improve the situation  
by collecting the most chemical and kinematical data of high quality available
in the literature and to investigate
the chemical properties of the thin and thick disks identified by the
simplest and most robust criterion, their velocity ellipsoids. 
Our aim is also to find the 
stellar parameters which allow the best separation of the thick 
disk and the thin disk for future investigations.

In order to study the properties of the thin disk and the thick disk separately, 
its is necessary to clearly identify stars of each population. In practice, this 
is not obvious because their distributions are
overlapping. The kinematical information is often used as a robust criterion for 
nearby stars but  
the deconvolution of velocity distributions is complexified by the fact that 
moving groups, superclusters and dynamical streams may
exist and translate
into inhomogeneities in the velocity ellipsoids. This problem has recently been 
revisited by Famaey et al. (2004) who identified several kinematical
subgroups in a local sample of giants, representative of the solar 
neighbourhood. Among them, the Hercules stream is
crucial because its motion, which is believed to have a dynamical origin, 
can be confused with that of the thick disk. Famaey et al. (2004) have 
estimated that 6.5\%  of their sample belongs this stream. If the hypothesis of a
dynamical origin is correct, then the Hercules stream must be made of a mixture of
stars of any population but mainly thin disk stars since they are more numerous.
We believe that most of the previous studies of the thick disk may have 
included Hercules stars which may have perturbed the results. In this study
we have carefully identified such stars.

We have focussed on the following  key points :
 (1) the compilation of all the useful 
stellar parameters with only very accurate measurements, (2) the compilation of
a large sample of disk stars with
a significant fraction of thick disk stars, 3) the identification of
pure thin disk and thick disk stars with a well defined robust criterion. 
The stellar parameters
useful for population studies and used in this paper are
the  velocities (distances, proper motions, radial velocities),
metallicities ([Fe/H]), abundance ratios and ages. The accuracy is 
guaranteed by
selecting only stars within 100\,pc of the Sun with a relative error on 
Hipparcos parallax lower than 10\% (ESA 1997), which have been  recently 
submitted to detailed abundance studies from high resolution, high signal 
to noise spectra.  In the last years,
several groups have produced studies of elemental abundances
involving 
a few dozen to hundreds of stars spanning a wide range of metallicity. 
We have combined them in order to build a large catalogue of
stars with known elemental abundances. The kinematical classification
of stars into the thin and thick disks is the most robust because
it is not model dependent and the kinematical properties of the two 
populations are well known. Moreover the combination of Hipparcos astrometric
measurements and radial velocities from echelle spectrographs makes the
3D velocities of nearby stars very accurate. We have assigned stars to the 
thin disk, 
the thick disk and the Hercules stream on the basis of their (U, V, W) velocity 
by computing their probability to belong to the corresponding velocity 
ellipsoid.

We describe in Sect. \ref{s:cat} the construction of a large
catalogue of elemental abundances from published data. Before combining
the data we have analysed the agreement of abundances between authors.  The 
distribution of abundances versus metallicity of the whole sample is shown. 
Sect. \ref{s:UVW} describes the age determination and
the kinematical data which was used to  compute (U,V,W) velocities and orbits.
The kinematical classification performed to select three subsamples
of stars representative of the thin disk, the thick disk 
and the Hercules stream is described in
Sect. \ref{s:class}. In Sect. \ref{s:trends} we report our findings on
 the chemical
properties of the three subsamples. Finally we discuss
the age distributions, the lack of a vertical  gradient in the thick disk 
and the case of several metal-rich outliers in Sect. \ref{s:grad}.

\section{The abundance catalogue}
\subsection{Construction of the catalogue}
\label{s:cat}
In order to build a large sample of elemental abundances, we
have compiled several studies from the literature
presenting
determinations of O, Na, Mg, Al, Si, Ca, Ti, Fe, Ni abundances. 
We have chosen in the literature the eleven most significant papers which
present large lists of stars (Table \ref{t:phys}).

\begin{table}[bht]
\centering
\caption[]{List of datasets included in the catalogue with the number
of stars in the chosen range of atmospheric parameters (see text) and 
the method
used for the Teff determination (ldr=line depth ratios) :
A04 = Allende Prieto et al. (2004), B03 = Bensby et al. (2003), B04a = Bensby et al.
(2004), C00 = Chen et al. (2000),
E93 = Edvardsson et al. (1993), F00 = Fulbright (2000), G03 = Gratton et al. (2003),
M04 = Mishenina et al. (2004), N97 = Nissen \& Schuster (1997), P00 = Prochaska et al. (2000),
R03 = Reddy et al. (2003)}
\label{t:phys}
{\scriptsize
\begin{center}
\begin{tabular}{lrll}
\hline
\hline
Reference   & N & Elements &Teff determination\\
\hline
A04  & 104 &Fe,Si,Ca,Mg,Ti,O,Ni       &b-y, B-V  \\
B03  &  66 &Fe,Si,Ca,Mg,Ti,Na,Al,Ni   &FeI\\
B04a &  66 &O                         &FeI\\
C00  &  90 &Fe,Si,Ca,Mg,Ti,O,Na,Al,Ni &b-y\\
E93  & 188 &Fe,Si,Ca,Mg,Ti,O,Na,Al,Ni &b-y\\
F00  & 100 &Fe,Si,Ca,Mg,Ti,Na,Al,Ni   &V-K, B-V, b-y\\
G03  & 116 &Fe,Si,Ca,Mg,Ti,O,Na,Ni    &B-V, b-y\\
M04  & 174 &Fe,Si,Mg,Ni	              &H$_\alpha$, ldr\\
N97  &  29 &Fe,Si,Ca,Mg,Ti,O          &FeI\\
P00  &  10 &Fe,Si,Ca,Mg,Ti,O,Na,Al,Ni &FeI\\
R03  & 181 &Fe,Si,Ca,Mg,Ti,Na,Al,Ni   &b-y\\
\hline
\end{tabular}
\end{center}
 }
\end{table}

Because
authors of spectral analyses do not use the same scales and methods,
systematics may
exist between  their results which cannot be combined without a
careful
analysis and eventually some kind of homogenisation. As most of the eleven
studies have stars in common, we were able to measure their agreement.
We have observed that disagreements were more frequent in the metal-poor
regime, and because we were interested in studying the disk population and
specially the interface between the thin disk and the thick disk, we have 
eliminated all stars with [Fe/H] $<$ -1.3. A larger 
dispersion observed  among cold stars convinced us to build our sample with
stars hotter than 4500K. The number of stars of the eleven studies which
fall within these limits is given in Table \ref{t:phys}. 

The effective temperature is the most critical
parameter in spectroscopic analyses and different methods for its
determination may lead to inhomogeneous scales. A systematic difference 
in temperature
translates into a systematic difference in metallicity thus in
abundances relative to iron. The eleven studies considered for
the catalogue have indeed used various methods, listed in Table \ref{t:phys}.
 Our first concern before combining
abundances from different authors was to quantify the agreement of
their temperature scales. We have used as reference temperatures the
determinations by Alonso et al. (1996), Blackwell \& Lynas-Gray (1998)
and di Benedetto (1998), which are known to be in good agreement in the
FGK regime. We have inspected for each paper, the agreement of
the temperature scale with the reference one.
Table \ref{t:teffABdiB} gives the mean differences and dispersions which
were obtained.
The dispersions are reasonable, ranging from 68K to 109K, typical of
the accuracy of 50-100K generally quoted in spectral analyses. The largest
offsets are observed for N97 and P00, however they correspond to the samples
with the poorest statistics. For the other studies, the offset is
lower than 75K. Such a difference is
not expected to strongly affect the abundances. According to M04,
 a difference of 100K in Teff translates into a difference of 0.07 dex in
[Fe/H] and a lower value for other elements.

\begin{table}[ht!]
\caption[]{Statistics of the Teff comparison 
to Alonso et al. (1996), Blackwell \& Lynas-Gray (1998) and di Benedetto (1998). 
}
\label{t:teffABdiB}
{\scriptsize
\begin{center}
\begin{tabular}{lrrr}
\hline
\hline
Ref   &   $\Delta_{\rm Teff}(K)$  & $\sigma_{\rm Teff}(K)$  &  N  \\
\hline
A04  & 16 &76 &43\\
B03  &-71 &72 &14\\
C00  &21  &109&18\\
E93  &-59 &68 &59\\
F00  &51  &85 &44\\
G03  &-15 &68 &42\\
M04  &-37 &77 &55\\
N97  &-90 &75 &8 \\
P00  &89  &68 &3  \\
R03  &23  &68 &11 \\
All  &-13 &88 &302 \\
\hline
\end{tabular}
\end{center}}
\end{table}

\begin{figure}[ht!]
\centering
\includegraphics[width=7cm]{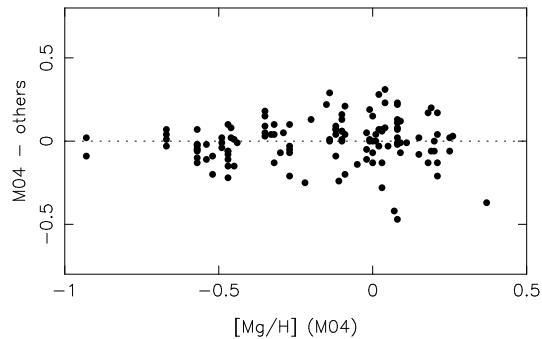} 
   \caption[]{Comparaison of [Mg/H] computed under NLTE approximation by M04 and LTE
   [Mg/H] computed by the other authors. No systematic trend can be observed. 
   The mean difference is null, the dispersion 0.13 dex for 126 stars in common.}
   \label{f:Mg}  
\end{figure}

\begin{figure}[ht!]
\centering
\includegraphics[width=7cm]{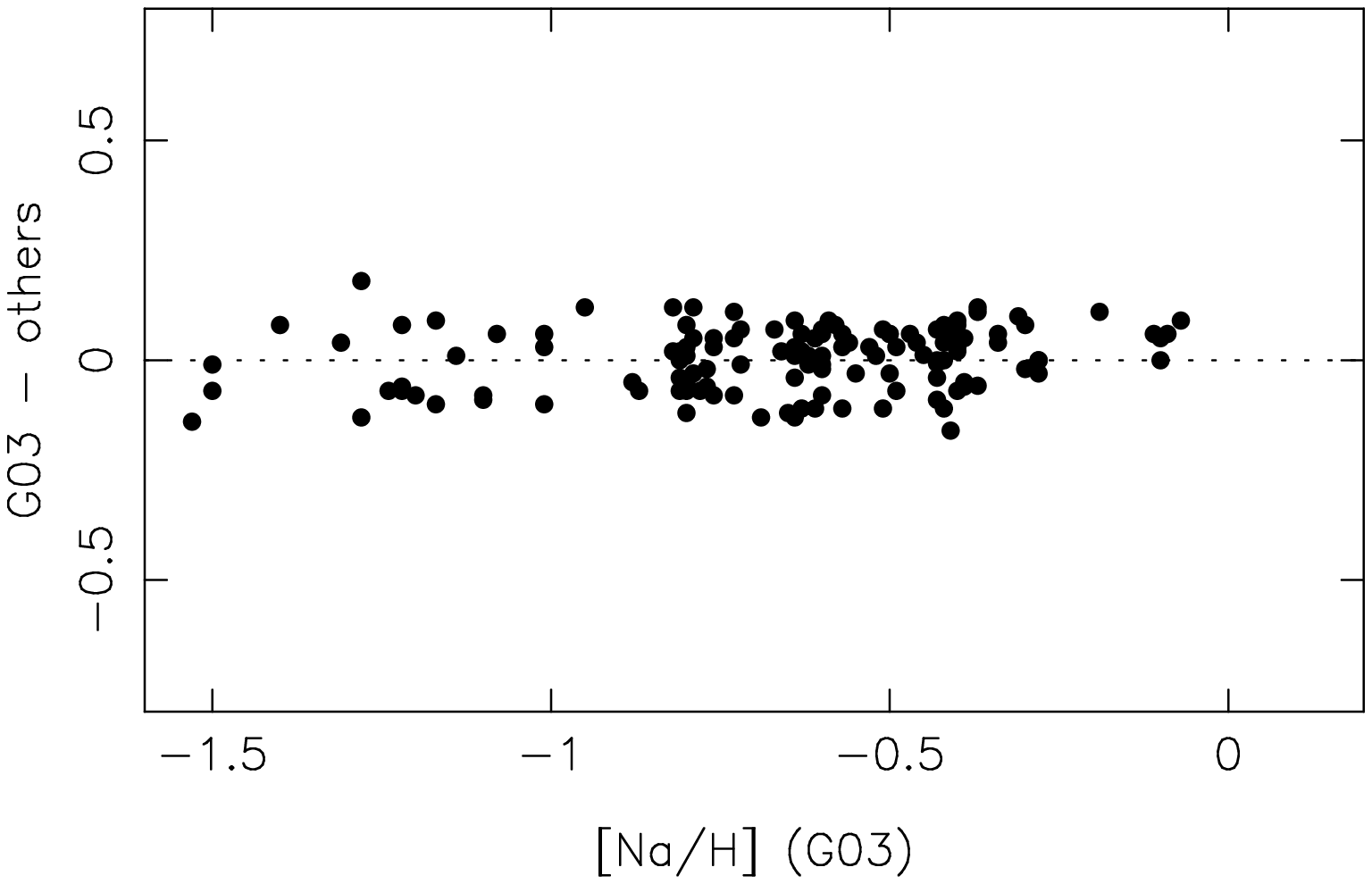} 
   \caption[]{Comparaison of [Na/H] computed under NLTE approximation by G03 and LTE
   [Na/H] computed by the other authors. No systematic trend can be observed.
   The mean difference is null, the dispersion 0.07 dex for 130 stars in common.}
   \label{f:Na}  
\end{figure}

Most of the determinations of elemental abundances considered here are
based on LTE spectral analyses. Two exceptions concern Na by
G03 and Mg by M04. Rigourously LTE and NLTE abundances should not
be combined because they rely on different physics and may lead to systematic
offsets depending on the spectral lines considered and the parameter range 
of the stars.
We have looked for such inconsistencies by comparing NLTE [Mg/H] by M04 and 
NLTE [Na/H] by G03 to their
LTE couterparts by the other authors for stars in common (Figs. \ref{f:Mg} and
\ref{f:Na}). The plots do not reveal
systematic effects.  For Mg, the mean difference is null with a dispersion 
of 0.13 dex for 126 stars in common. For Na, the mean difference is also null 
with a dispersion 0.07 dex for 130 stars in common. The lack of an offset
between these abundances led us to keep Na and Mg NLTE determinations in the
catalogue in order to enlarge our sample. 

\begin{figure}[ht!]
\centering
\includegraphics[width=7cm]{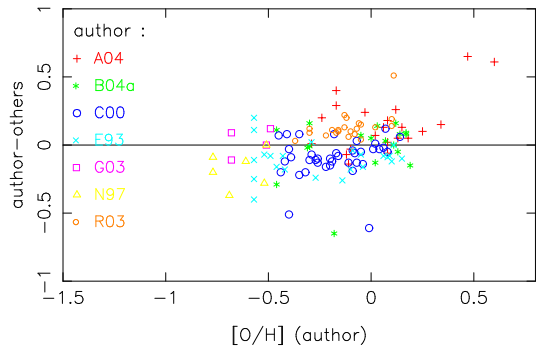} 
   \caption[]{Differences between [O/H] on the LTE 6300\AA\, 
 scale derived by authors having stars in common.}
   \label{f:Ox}  
\end{figure}

The case of oxygen is more complex.
Oxygen abundances can be determined either from the forbidden [OI]
lines at 6300\AA\, and 6363\AA\, or from the IR triplet OI lines at 7774\AA\,.
The lines at 7774\AA\, are strong and clean but affected by NLTE effects, whereas
lines at 6300\AA\, and 6363\AA\, are weak and blended but unaffected by NLTE 
effects. According to the IAU recommendations, the latest lines are the best indicators.
We have thus selected in the considered papers
only oxygen abundances on the LTE 6300\AA\, scale, obtained either directly or with a correction.
E93 have calibrated a correction to pass from  the LTE oxygen abundances  derived with 
the 7774\AA\, line to those derived with the 6300\AA\, line. C00 have used this transformation, and we
have also used it to correct the LTE 7774\AA\, determinations by  N97 and P00. 
A04 and B04a have used directly  the 6300\AA\, lines.
Finally G03 have provided LTE 6300\AA\, abundances for 22 stars and NLTE 7774\AA\, determinations
for 68 stars. We have only kept the 22 LTE ones.
Fig. \ref{f:Ox} shows the typical dispersion that affects oxygen abundances determined by
different authors, even if they are on the same scale. The standard deviation of the plotted distribution
is 0.19 dex (132 values, 0.14 dex when outliers are removed). 
There are 3 outliers : HD172051 with [O/H]=+0.47 (A04) and 
[O/H]=-0.18 (B04a), HD109303 with [O/H]=+0.11 (R03) and [O/H]=-0.40 (C00),  HD210027 with
[O/H]=+0.60 (A04) and [O/H]=-0.01 (C00). A04 determinations are systematically higher 
than the others by 0.11 dex and several stars have
unexpectedly high values of [O/H] (HD157214 and HD144579 for instance). Considering that the total 
dispersion of differences between authors decreases to 0.13 dex (85 values) without A04, we have
not considered the oxygen abundances by A04 in the final catalogue. This is beyond the scope of this
paper to analyse why A04 doesn't seem to be on the same [O/H] scale than the other authors.


We have continued our verification by
comparing [Fe/H] and abundance ratios of the different studies 
having at least 10 stars in common. 
We have observed the largest systematic differences, ranging from 0.08 to 0.10 dex,
for [Ti/Fe] and [Mg/Fe] (A04 vs E93), and for [Mg/Fe] (A04 vs C00).
The median offset between the studies which could be compared 
is 0.02. The largest dispersions concern E93 vs C00 for [Ti/Fe] and
[Al/Fe] (0.10 dex), and M04 vs A04 for [Fe/H] (0.11 dex). The median dispersion 
is 0.045 dex (a 3$\sigma$ rejection of a few outliers was performed to compute 
these statistics). These values 
are reasonable, considering realistic error bars 0.10 dex on [Fe/H] and
0.06 dex on abundance ratios. We have thus constructed our catalogue of abundances
by combining the results of the eleven selected papers,
adopting a simple average when several determinations were available for
a given star. The
final catalogue consists of 743 stars, only available in electronic form
at the CDS, Strasbourg. Table \ref{t:count_ab} gives the
number of stars available per element. 

\begin{table}[ht!]
\caption[]{For each element, number of stars with a determination of
abundance}
\label{t:count_ab}
\begin{center}
\begin{tabular}{ccccccccc}
\hline
\hline
Fe & O & Na & Mg & Al & Si & Ca & Ti & Ni \\
\hline
743 & 415 & 568 & 725 & 509 & 743 & 641 & 630 & 739 \\ 
\hline
\end{tabular}
\end{center}
\end{table}

\subsection{Abundance trends in the whole sample}

\begin{figure*}[]
\centering
\includegraphics[width=7cm]{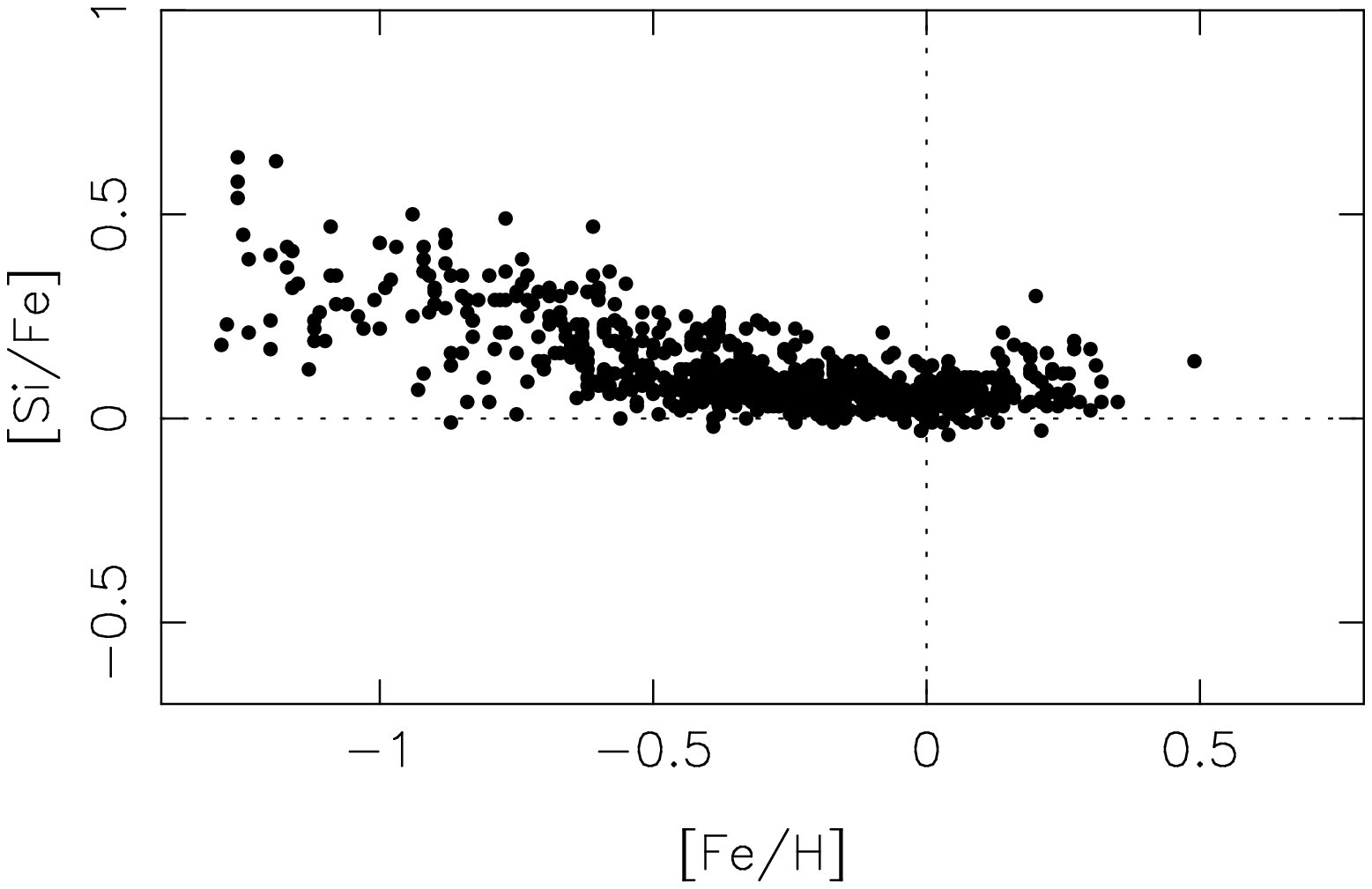}
\includegraphics[width=7cm]{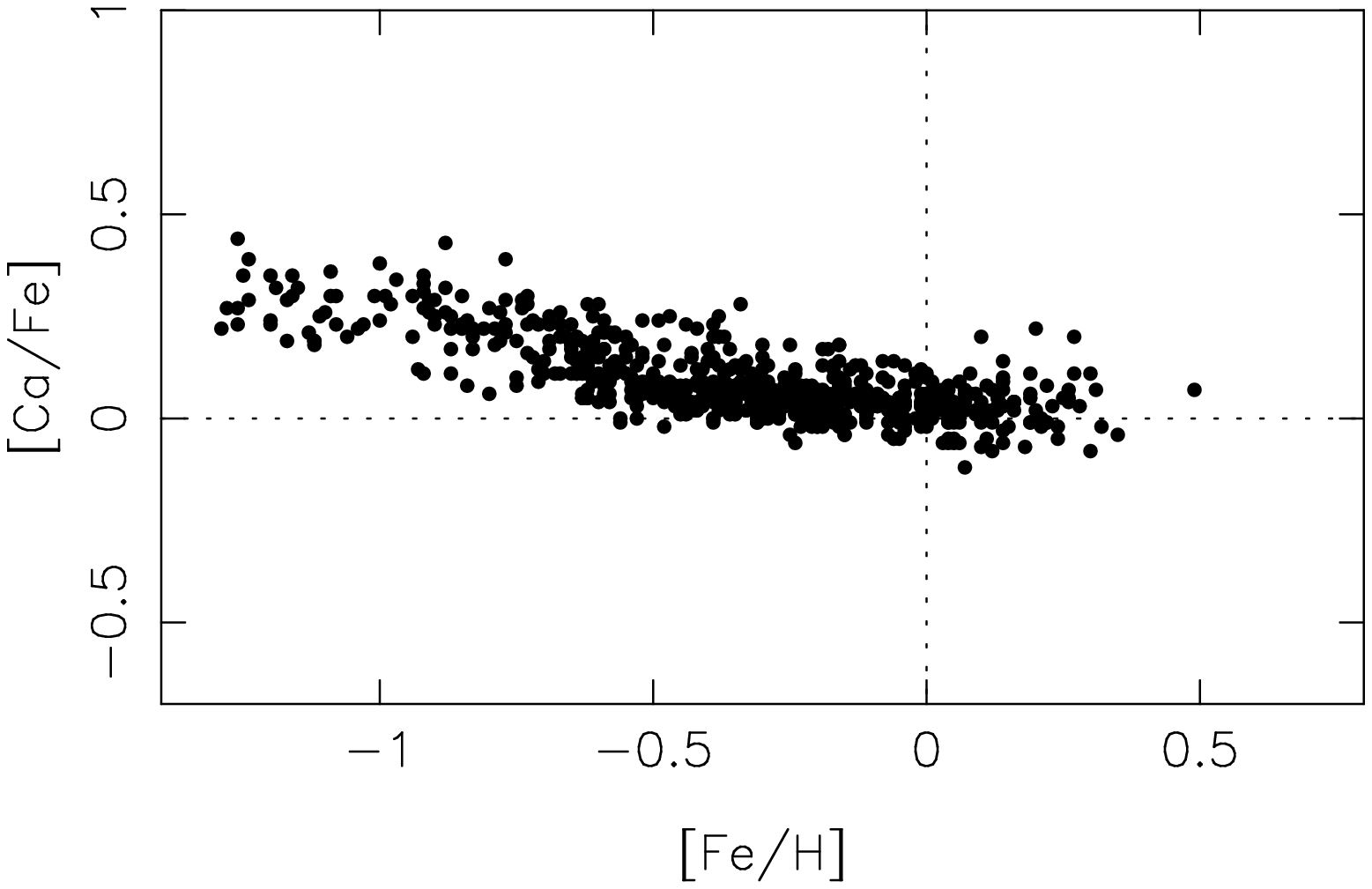}
\includegraphics[width=7cm]{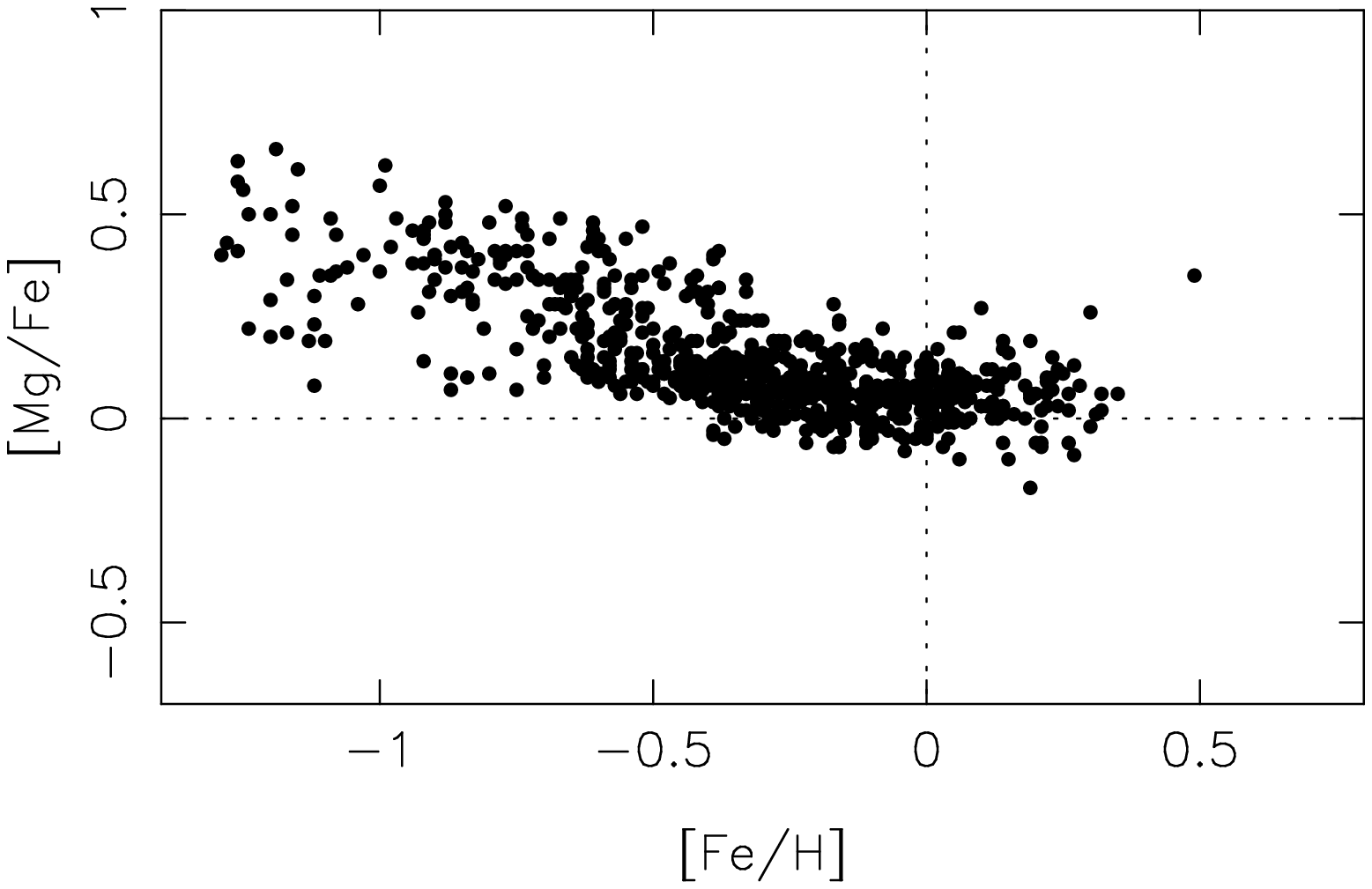}
\includegraphics[width=7cm]{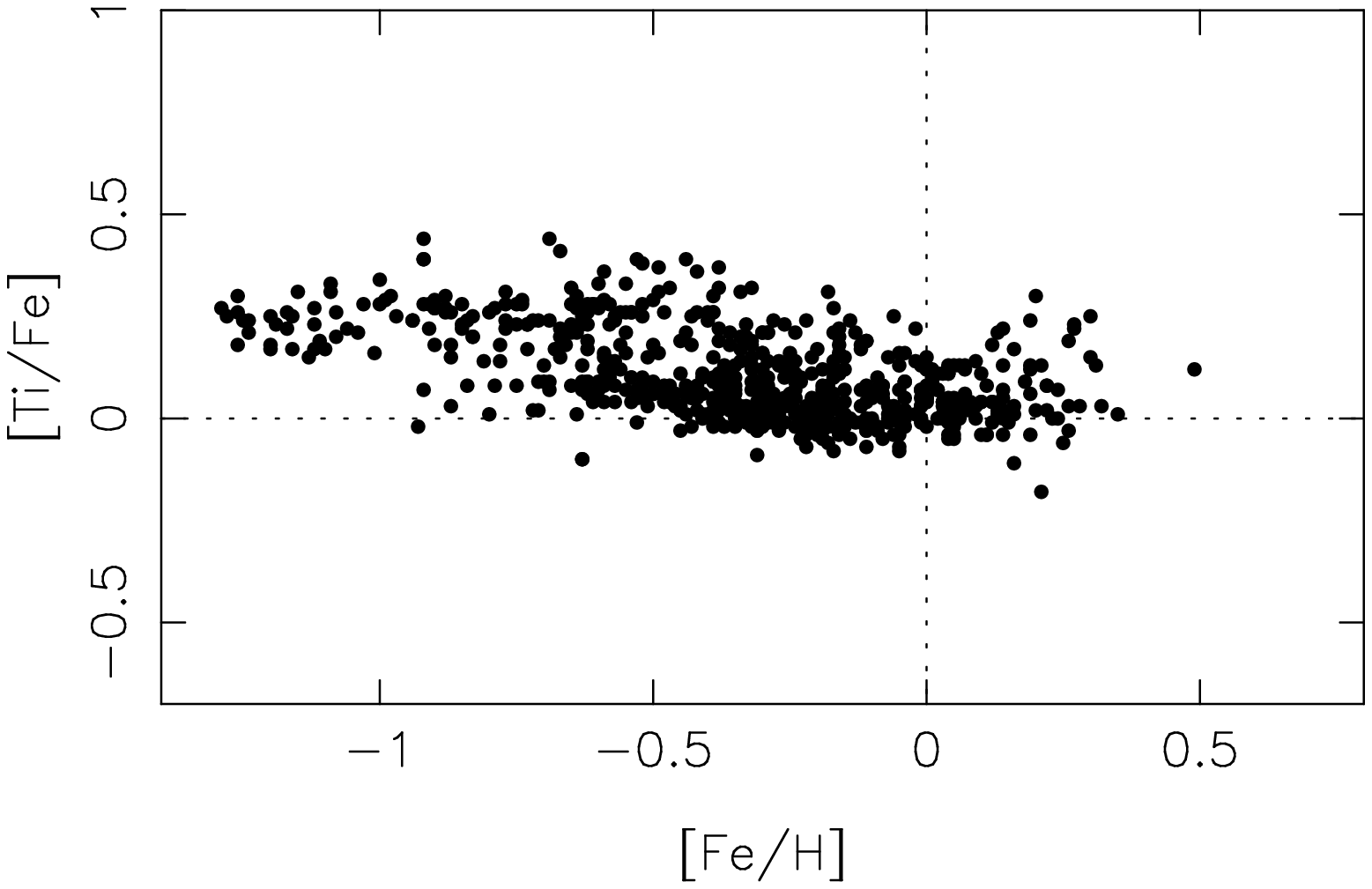}
\includegraphics[width=7cm]{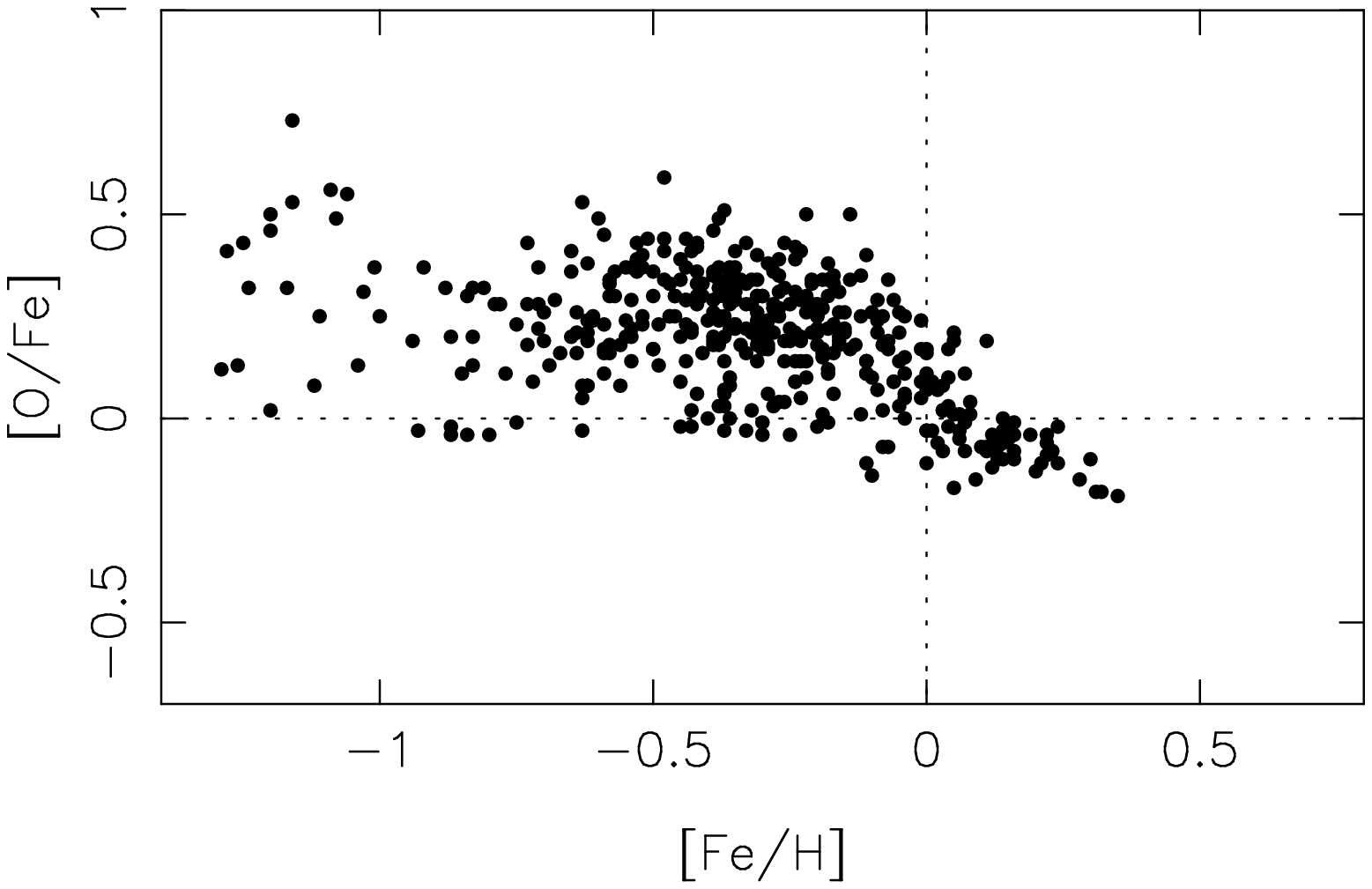}
\includegraphics[width=7cm]{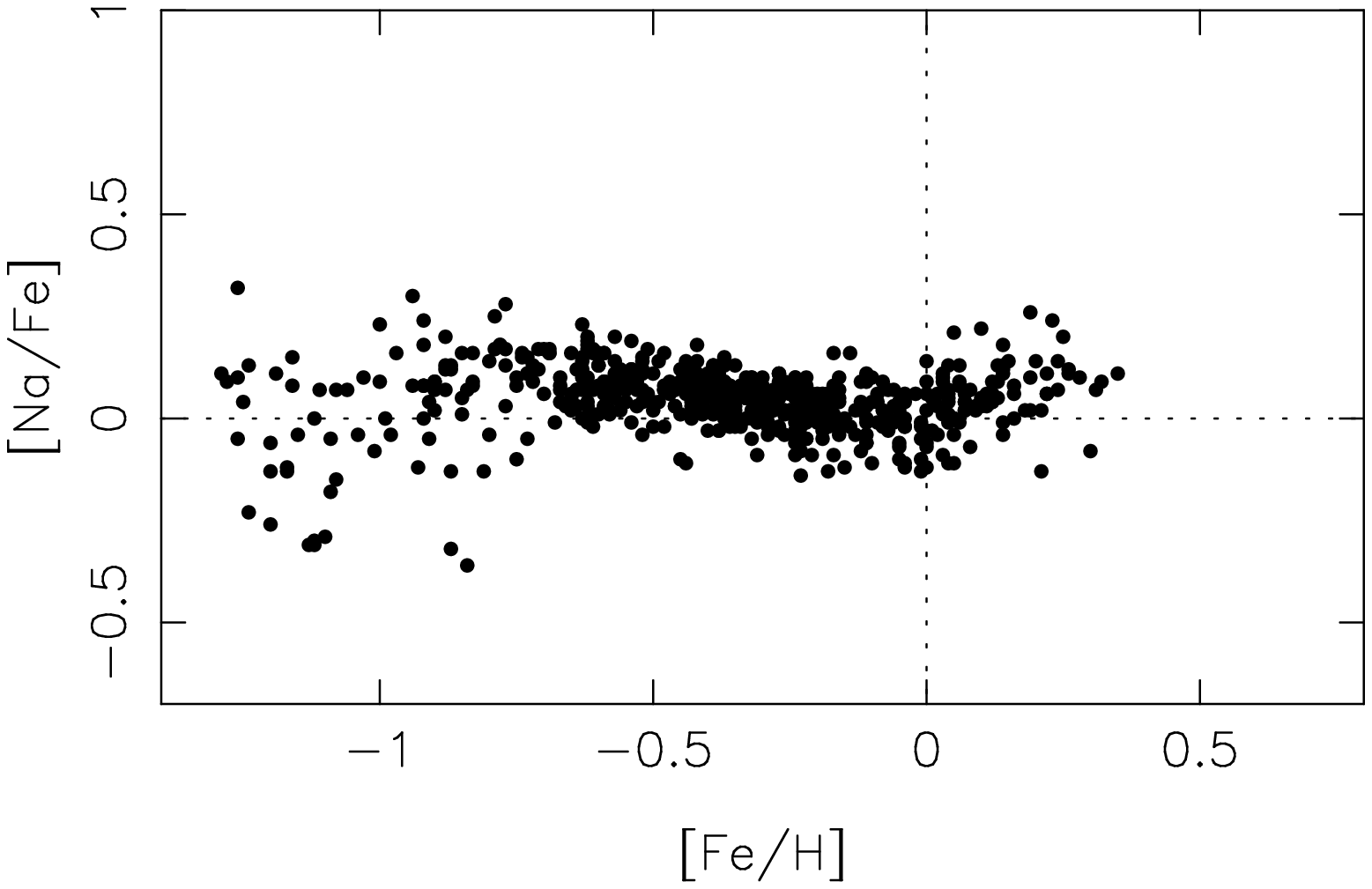}
\includegraphics[width=7cm]{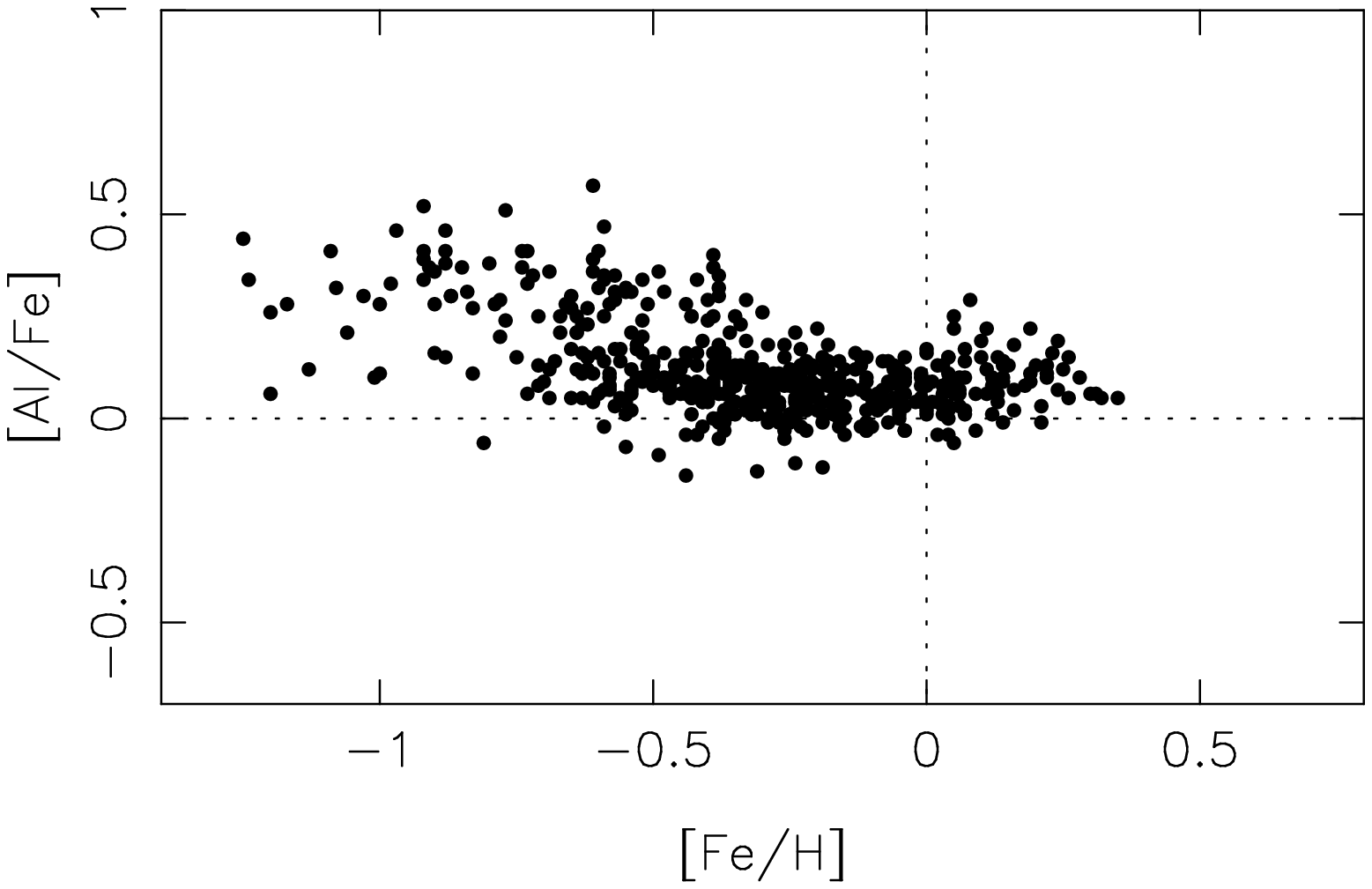}
\includegraphics[width=7cm]{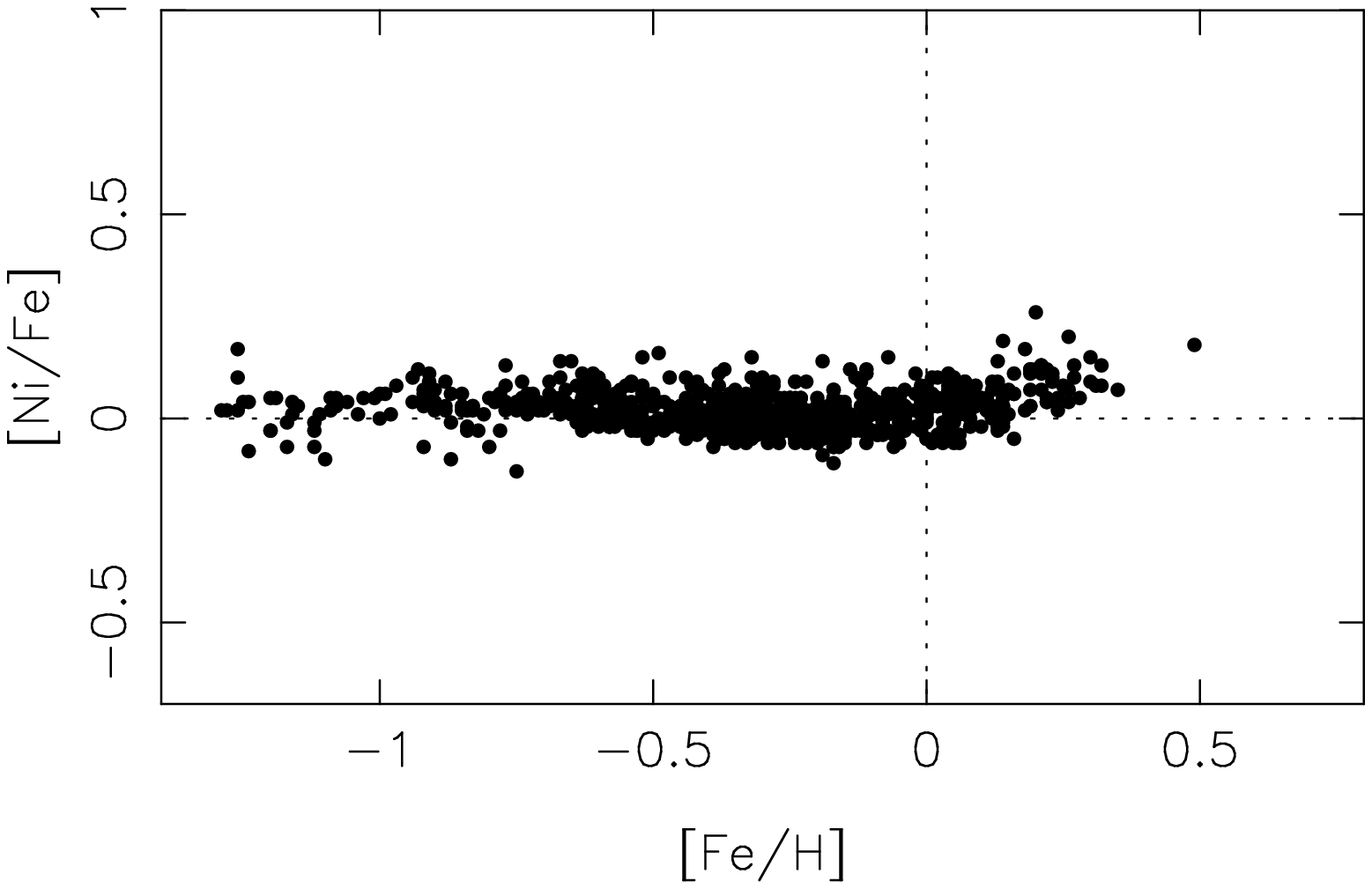}
\caption[]{[X/Fe] vs [Fe/H]}
\label{f:xfe}
\end{figure*}

For each element X, Fig. \ref{f:xfe} represents the distribution of the
sample in the plane [X/Fe] vs [Fe/H]. 

As expected, $\alpha$ elements (Mg, Si, Ti and Ca) show a similar behaviour with a 
decreasing trend as metallicity increases, the distribution being flat at solar 
metallicity.
 There are however some differences between these elements : Mg 
presents a higher dispersion at low metallicity, contrary to Ca which deacreases more
regularly.  Oxygen,
which is also an $\alpha$ element continues to decrease in the metal rich regime
and exhibits a large spread. The large dispersion of [O/Fe] might be cosmic, but also due to 
measurement errors and inhomogeneities in the studies which were combined. 

Interestingly, it seems
that Al has the same behaviour as $\alpha$ elements, a similarity which
has already been reported by E93 and B03. The two main
features concerning Na are a change of dispersion at [Fe/H]$\simeq$-0.70,
the metal-rich part showing a very low dispersion, and a rise at super-solar
metallicities.
The remarkable low dispersion of [Ni/Fe] in the whole range of metallicity, measured
to be 0.04 dex
implies (1) that the different studies of our compilation are in excellent agreement 
for this element, (2) that the cosmic scatter is very low. As an iron 
peak element, Ni follows very well the metallicity.
Most elements, except oxygen, show a rise of enhancement with respect to iron at [Fe/H]$>$0.
Oxygen is the only element which is depleted in the metal rich regime. According to current
models of chemical evolution of the Galaxy, $\alpha$ elements like O and Mg have the same source
of production, hence should have the same trends. This is not observed in the metal rich regime
but an interpretation falls outside the scope of this paper. 

The Sun appears to be slightly deficient in all elements compared
to stars with similar metallicity. The peculiarity of the solar abundance ratios was
previously noticed by A04 and E93 with a tentative explaination related to its larger 
age.

\section{Velocities, orbits and ages}
\label{s:UVW}
In order to compute accurate velocities (U,V,W) we have first 
cross-correlated the abundance catalogue with Hipparcos, selecting stars
with 
$\pi >$10 mas and ${\sigma_{\pi}\over \pi} <$0.10.  Then we have searched 
radial velocities in the following sources : Prugniel \& Soubiran (2001), 
Nidever et al. (2002), Nordstr\"om et
al. (2004), Gratton et al. (2003), Barbier-Bossat et al. (2000). 
Positions, distances, proper motions and radial velocities have 
been combined
to compute the 3 components (U,V,W) of the spatial velocities with
respect to the Sun. For most stars the uncertainty on the components of 
the spatial velocity is lower than 1 \kms.

 We have been tempted to adopt the ages derived by Nordstr\"om et al. (2004) for
447 stars that we have in common with this catalogue. However their isochrone ages
have been computed with photometric metallicities, of much lower
accuracy than the spectroscopic ones that are available for our stars. Instead 
the derivation of ages was very kindly done for us by Fr\'ed\'eric Pont making use
of the Bayesian method of Pont 
\& Eyer (2004) together with the high quality data (Hipparcos luminosities and spectroscopic temperatures 
and metallicities) that we have compiled and the theoretical isochrones from Padova (Girardi et al. 2000). 
Unfortunatly the current version of the software does not use $\alpha$-enhanced isochrones. On average
the $\alpha$-enhancement of the most metal-poor stars of our sample is [$\alpha$/Fe]=+0.3. Bensby et al 
(2004b) have shown how $\alpha$-enhancement changes isochrone ages. In their example at [Fe/H]=-0.70 
(see their Fig. 3), the shift between the isochrones at [$\alpha$/Fe]=+0.3 and [$\alpha$/Fe]=0.0 may reach
5 Gyr in the evolved parts. Ages may be overestimated when the $\alpha$-enhancement is not taken into 
account. However it can be noticed in their Fig. 3 that the difference between enhanced and non-enhanced
isochrones is equivalent at the first order to a temperature shift. As explained in Pont \& Eyer (2004), 
the temperature match
between observations and models is one of the key point of the method. The temperature shift which is
performed to make the model match the observations at a given metallicity corrects also the effects due
to the neglected [$\alpha$/Fe] ratio.
With the Bayesian  method the most probable age is
estimated together with a probability distribution function which was used
to define a subsample of well defined ages.  Ages have been computed for all
stars, but then we have eliminated those corresponding to
stars showing a wide or asymetrical age probability distribution function,
or with a position in the HR diagram far from the isochrones
of same metallicity. With those restrictions we have good estimations
of ages for 322 stars.

The orbital parameters have been computed by integrating the
equations of motion in the
galactic model of Allen \& Santillan (\cite{allen}). When available we
have adopted the age of the star as the integration time, otherwise a default
value of 5 Gyr was adopted. The adopted velocity of the Sun with respect 
to the LSR is (9.7, 5.2, 6.7)
\kms (Bienaym\'e 1999, U positive towards the galactic center), the solar galactocentric 
distance  ${\mathrm R}_{\odot}=8.5$ kpc
and circular velocity ${\mathrm V_{lsr}}=220$ \kms. In order to focus
on the populations of the disk, we have eliminated from the sample
several stars with halo kinematics. The final sample with abundances, 
velocities and orbits  includes 639 stars. 

\begin{figure}[hbtp]
\includegraphics[width=7cm]{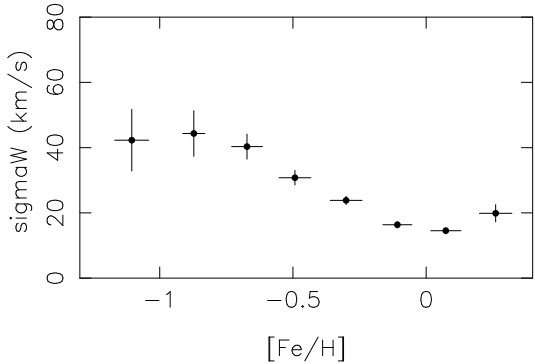}
\caption[]{$\sigma_W$ vs [Fe/H] of the whole sample in several bins of metallicity. Error
bars correspond to the standard error for $\sigma_W (\sigma_W/\sqrt{2N})$ and 
standard deviation for [Fe/H].}
\label{f:feh_W}
\end{figure}

\begin{figure}[hbtp]
\includegraphics[width=7cm]{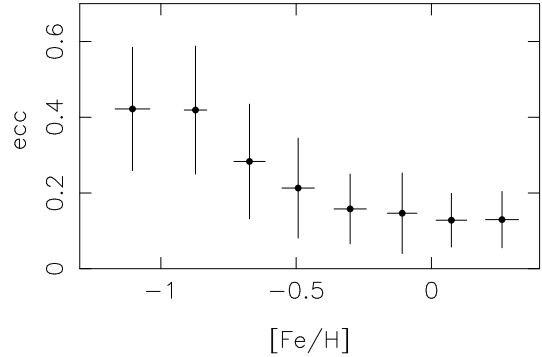}
\caption[]{Eccentricity of the orbits vs [Fe/H] of the whole sample in several 
bins of metallicity. Error
bars correspond to the standard deviations.}
\label{f:feh_ecc}
\end{figure}

\begin{figure}[hbtp]
\includegraphics[width=7cm]{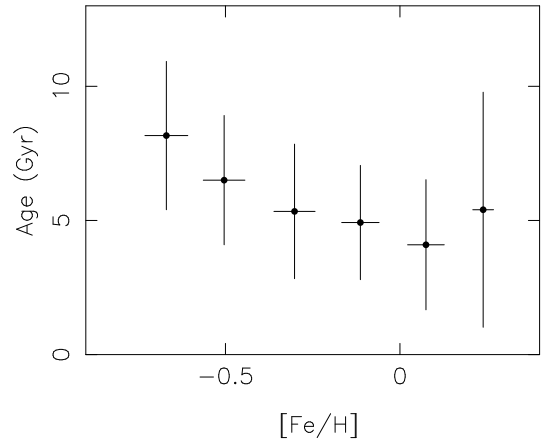}
\caption[]{Age vs [Fe/H] in several bins of metallicity for stars having 
well defined ages. Error
bars correspond to the standard deviations.}
\label{f:feh_age_tot}
\end{figure}

Figs. \ref{f:feh_W} and \ref{f:feh_ecc} present 
the dispersion of the vertical velocities W and the eccentricities of the orbits
 in several bins of metallicity. Three regimes can be distinguished.
The three bins of highest metallicity show a nearly flat distribution with
$\sigma_W \simeq 15$ \kms and $ecc \simeq 0.09$ showing that this metallicity 
interval is dominated by the thin disk. Interestingly the bin at [Fe/H]=+0.26
has $\sigma_W =20$ \kms suggesting that there are super metal rich stars with
hotter vertical kinematics. The increase of $\sigma_W$ and
$ecc$ as metallicity decreases in the next three bins reflects the mixture
of the thin disk and the thick disk, with a growing number of thick disk stars.
The last two bins of lowest
metallicity saturate at $\sigma_W \simeq 45$ \kms and $ecc \simeq 0.45$.
A smooth AMR is visible in Fig. \ref{f:feh_age_tot}. Several old super metal-rich 
stars are responsible of the jump and large dispersion in the last bin of
metallicity. The same stars have also hot vertical kinematics.
 
\section{Kinematical classification}
\label{s:class}
In order to investigate the abundance trends in the thin disk and the thick
disk separately, we have classified the stars into these
populations, using the kinematical information. We have performed such a 
deconvolution
in previous studies (Soubiran et al. 2003 and M04) where
we have noticed that metal rich stars with high eccentricity but low scale
height were assigned to the thick disk whereas they could have another origin.
The recent study by Famaey et al. (2004) has confirmed that 6.5\% of the stars
in the solar neigbourhood belong to the Hercules stream, which has a mean
rotational motion with respect to the Sun very similar to that of the thick
disk $(V=-51$ \kms) with a significant radial motion of $U=-42$ \kms. 
 Famaey et al. (2004) have identified a total of six kinematical structures in their
sample of nearby giants : the background stars corresponding to the mixed population
of the thin disk, the high velocity group corresponding to the thick disk, the Hercules stream,
the young group, the Hyades-Pleiades supercluster and the Sirius moving group. The three
later groups correspond to young stars with kinematics typical of the thin disk. As can
be seen in their Fig. 9, these groups of young stars appear as concentrations inside 
the velocity ellipsoid of the thin disk. They represent a peculiar class of thin
disk stars and were not considered separately in this study. On the contrary the 
Hercules stream 
is of high interest for our probe of the thin disk to thick disk interface since 
its kinematics is
exactly intermediate between to two populations. 
The authors
explain that this stream has a dynamical origin. Contrary to a moving
group which involves stars which were born at the same 
place and at the same time, a dynamical stream affects stars of
any age and population. The effect of the bar in the 
central parts of the Galaxy is to move stars onto excentric orbits which tend then to
mimic thick disk stars. Such stars, mainly thin disk stars with
perturbed kinematics, could have polluted previous samples
of thick disk stars selected on kinematical criteria and thus must be taken 
into account.
The kinematical classification that we have performed is based on  the 
assumption that 
our whole catalogue is dominated by three 
populations, the thin disk, the thick disk and the Hercules stream, all having
gaussian velocity distributions.

We have  computed for every star 
its probability to
belong to each of the three populations on the basis of its (U,V,W) velocity
and the gaussian velocity ellispoid of the corresponding population (equations
can be found in M04). The
kinematical parameters (mean, standard deviation, vertex deviation) which
define the velocity ellipsoids, as well as the proportions of the three 
populations in the sample,
have to be known.  We adopt for the kinematical parameters those determined 
by Soubiran et al. (2003) 
for the old thin disk and the thick disk and by Famaey et al. (2004) for the
Hercules stream (Table \ref{t:param_kine}). The fraction of each population in
the sample is more difficult to evaluate.
We are aware that our whole sample might be biased in favour of high velocity and 
metal poor stars due to the subject
of the eleven studies we are based on. Consequently the
proportion of thick disk and Hercules stars is expected to be higher 
than in the
complete sample of Famaey et al. (2004) which is more representative of the content
of the solar neighbourhood. In a similar way as
described in M04 we have applied to our sample of (U,V,W) a non-informative
algorithm
of deconvolution of gaussian distributions which showed that 72\% of the stars 
have kinematics typical of the thin disk and 28\% typical of the thick disk
or the Hercules stream. Adopting the same ratio of thick disk
to Hercules stream as Famaey et al. (2004), we have considered that 19\% of our
sample correspond to the thick disk, 9\% to the Hercules stream.

\begin{table}[bht]
\centering
\caption[]{Kinematical parameters of the three 
considered groups and their proportion in our sample.} 
\label{t:param_kine}
{\scriptsize
\begin{center}
\begin{tabular}{lrrr}
\hline
\hline
 & thin disk & thick disk & Hercules \\
\hline 
p (\%)            & 72  & 19& 9 \\
U (\kms)          & 0   & 0 & -42 \\
V  (\kms)         &-12  &-51& -51 \\
W  (\kms)         &  0  & 0 & -7 \\
$\sigma_U$ (\kms) & 39  &63 & 26 \\
$\sigma_V$ (\kms) & 20  &39 & 9 \\
$\sigma_W$(\kms)  & 20  &39 &17 \\
$l_v (\deg)$&  0  & 0 &-5.7 \\
\hline
\end{tabular}
\end{center}}
\end{table}

\begin{figure}[hbtp]
\includegraphics[width=7cm]{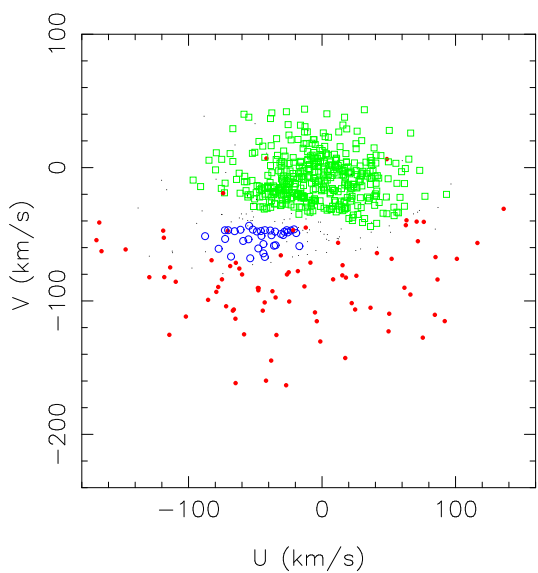}
\caption[]{The sample of 639 stars plotted in the (U,V) plane. Thick disk stars are
represented by red dots, thin disk stars by open green squares and Hercules stars
by open blue circles. The small dots correspond to stars which could not be
classified into the 3 subgroups. 
}
\label{f:UV}
\end{figure}

\begin{figure}[hbtp]
\includegraphics[width=7cm]{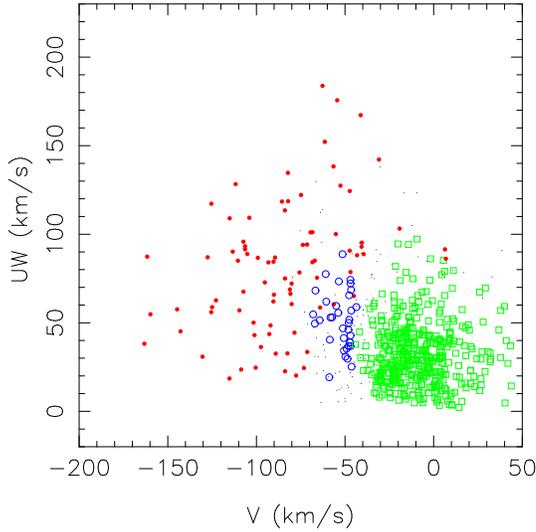}
\caption[]{Like Fig. \ref{f:UV}, but in the Toomre diagram : $UW=\sqrt{U^2+W^2}$ \kms}
\label{f:toom}
\end{figure}

According to the computed probabilities,
we have selected three subsamples representative of the thin disk, the thick disk and
the Hercules stream. The thin disk and the thick disk samples include respectively
428 and 84 stars having a probability higher than 80\% to belong to these
populations.  This probability cut was chosen as a compromise between having a sufficient
number of thick disk stars and avoiding the contamination of the sample with stars with 
intermediate kinematics. It is more difficult to isolate Hercules stars since their velocity 
distribution is greatly overlaping that of the thin and thick disks. Consequently there are only
5 stars with a probability higher than 80\% to belong to the Hercules stream. We had to lower the
probability limit down to 50\% to obtain 44 stars for the Hercules sample. Only 81 stars could
not be classified into one of the three groups. These remaining stars represent a mixture
of intermediate stars which are not considered in the following since our aim to work on samples
as pure as possible. The (U,V) plane and Toomre diagram of the whole sample is shown
in Fig. \ref{f:UV} and Fig. \ref{f:toom}, the three subsamples being highlighted in 
different colours and symbols. The Hercules stream is clearly visible
as a concentration between the two disks.

 How does our kinematical classification compare with those previously
performed in other works ? A04, E93, F00, G03 and N97 have not considered the
thick disk specifically. P00 has selected 10 thick disk stars 
with -1.1 $\le$ [Fe/H] $\le$
-0.4 and Zmax $>$ 600 pc and -20 $\le$ V$_{LSR}$ $\le$ -100 \kms. Only 5 of them 
remain in our catalogue due to our restrictions on distances, all 
of them classified into the thick disk. If we apply P00's criteria to our catalogue, we
only select 22 thick disk stars. These criteria are thus very efficient to avoid 
intermediate or Hercules stars but they are far too restrictive to construct a large
thick disk sample.
C00 have proposed their stars with [Fe/H] $\le$ -0.6
and V $\le$ -40 \kms to be thick disk stars. The same criterion applied to
our catalogue leads to a mixed sample including 58 thick disk stars,
13 Hercules stars, 1 thin disk star and 12 intermediate stars.  
R03 have performed a more complicated selection of thick
disk stars involving
metallicity, mean of peri- and apogalactic distances R$_m$, V velocity and age.
Doing the same without the age 
restriction (age $>$ 10 Gyr), we select 25 thick disk stars and 4 Hercules stars. Taking
the age restriction into account, only 3 thick disk stars are selected. These results
show that a restriction on  Zmax is more efficient than on V or R$_m$ to isolate pure
thick disk stars. Nevertheless a restriction on [Fe/H] is incompatible with the
study of the thick disk's  metallicity distribution and abundance trends.
The kinematical classification 
performed by B03 is the only one which is quite similar to ours in its principle
although they have used probability ratios. 
Among their 21 thick disk stars, only 2 of them (HD210483 and HD212231) are not
part of our thick disk sample because of our high probability cut (they
have a probability of 78\% and 75\% respectively to belong to the thick disk
with our classification). Among their 45 thin disk stars, HD003735 has a high
probability to belong to the Hercules stream (73\%) while HD210277 is
intermediate between the thin disk (probability 55\%) and the Hercules stream 
(probability 31\%). The classification by M04 is strictly the same as ours, but without
the Hercules stream. Among their 30 thick disk stars, 5 
have moved into the Hercules stream, and 5 are transition stars which 
could not be classified.

The catalogue including abundances, kinematical and orbital parameters, ages, membership probabilities
is available at the CDS, in the form of a single table. For each star it contains the following 
information (when available) : 
Column 1 lists the name of the source, Column 2 gives the metallicity [Fe/H], Column 3 to 10 give
the abundances relative to iron of O, Na, Mg, Al, Si, Ca, Ti and Ni, Columns 11 to 13 give the three 
components (U,V,W)
of the heliocentric galactic velocity, Columns 14 to 17 give the orbital parameters Rmin, Rmax, Zmax, 
eccentricity, Columns 18 and 19 give the age and its relative error, Columns 20 to 22 give the membership 
probability of the star to belong to the thin disk, the thick disk and the Hercules stream respectively.

\section{Abundance trends}
\label{s:trends}

\begin{figure*}[]
\centering
\includegraphics[width=7cm]{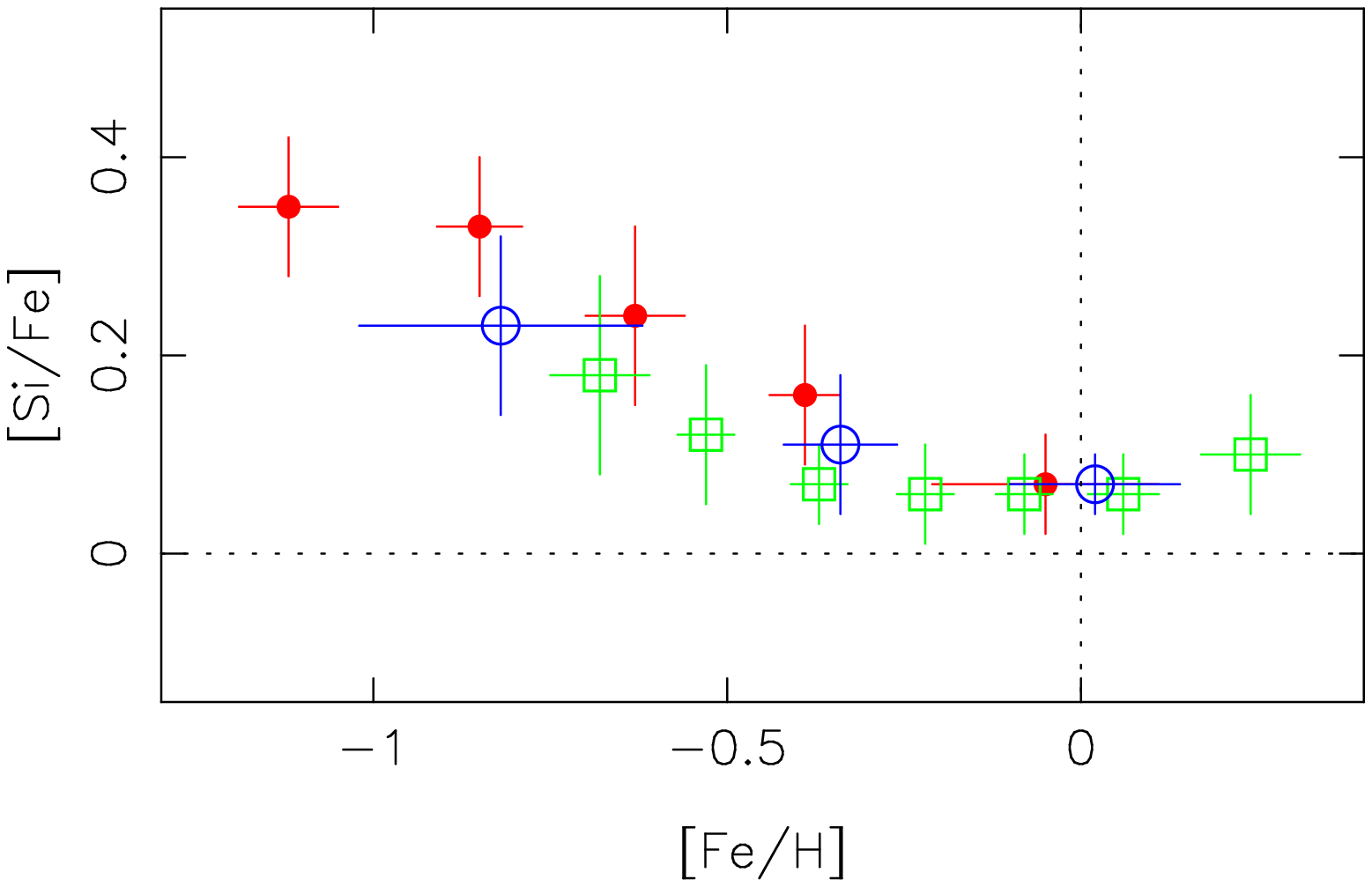}
\includegraphics[width=7cm]{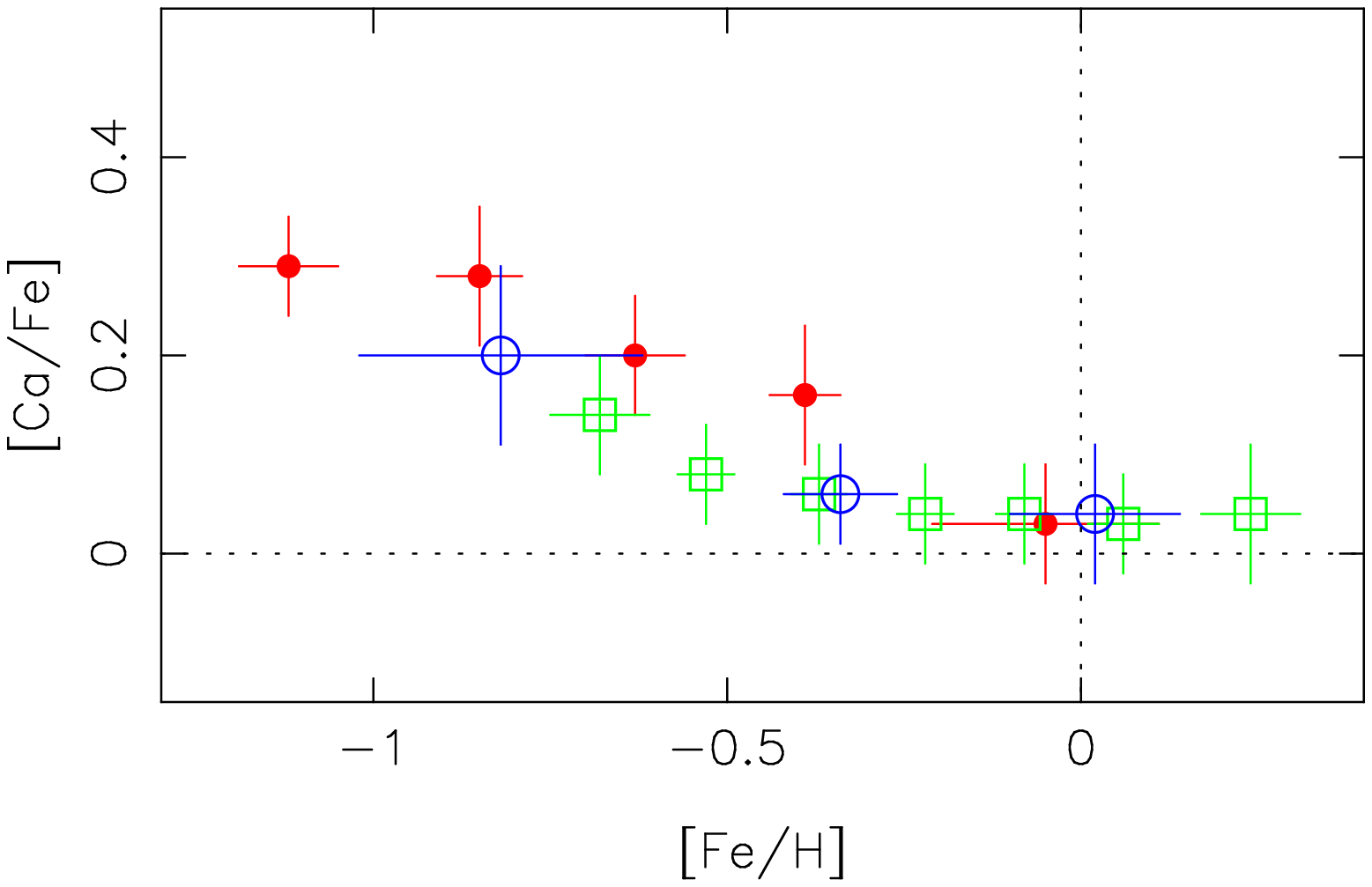}
\includegraphics[width=7cm]{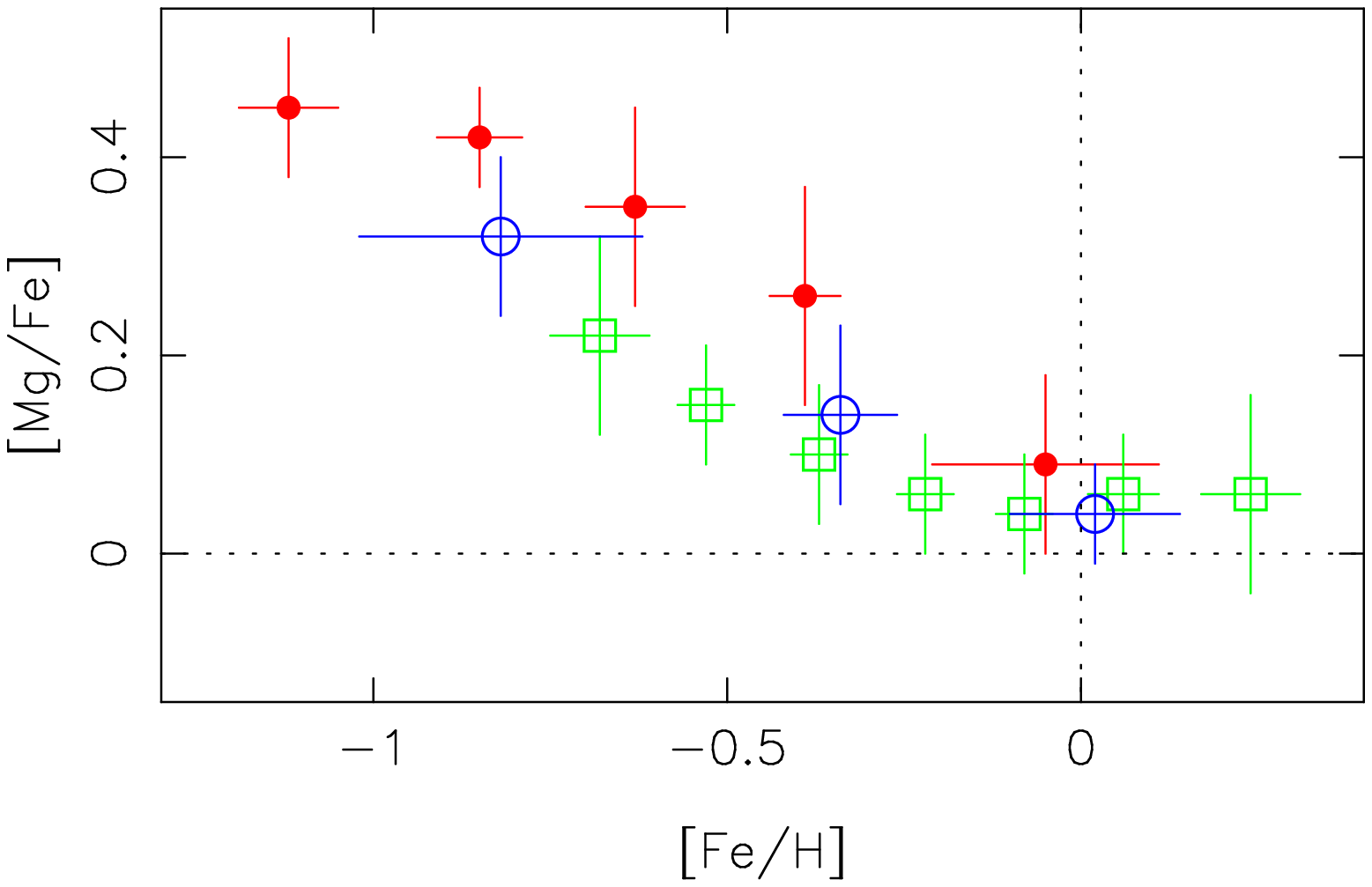}
\includegraphics[width=7cm]{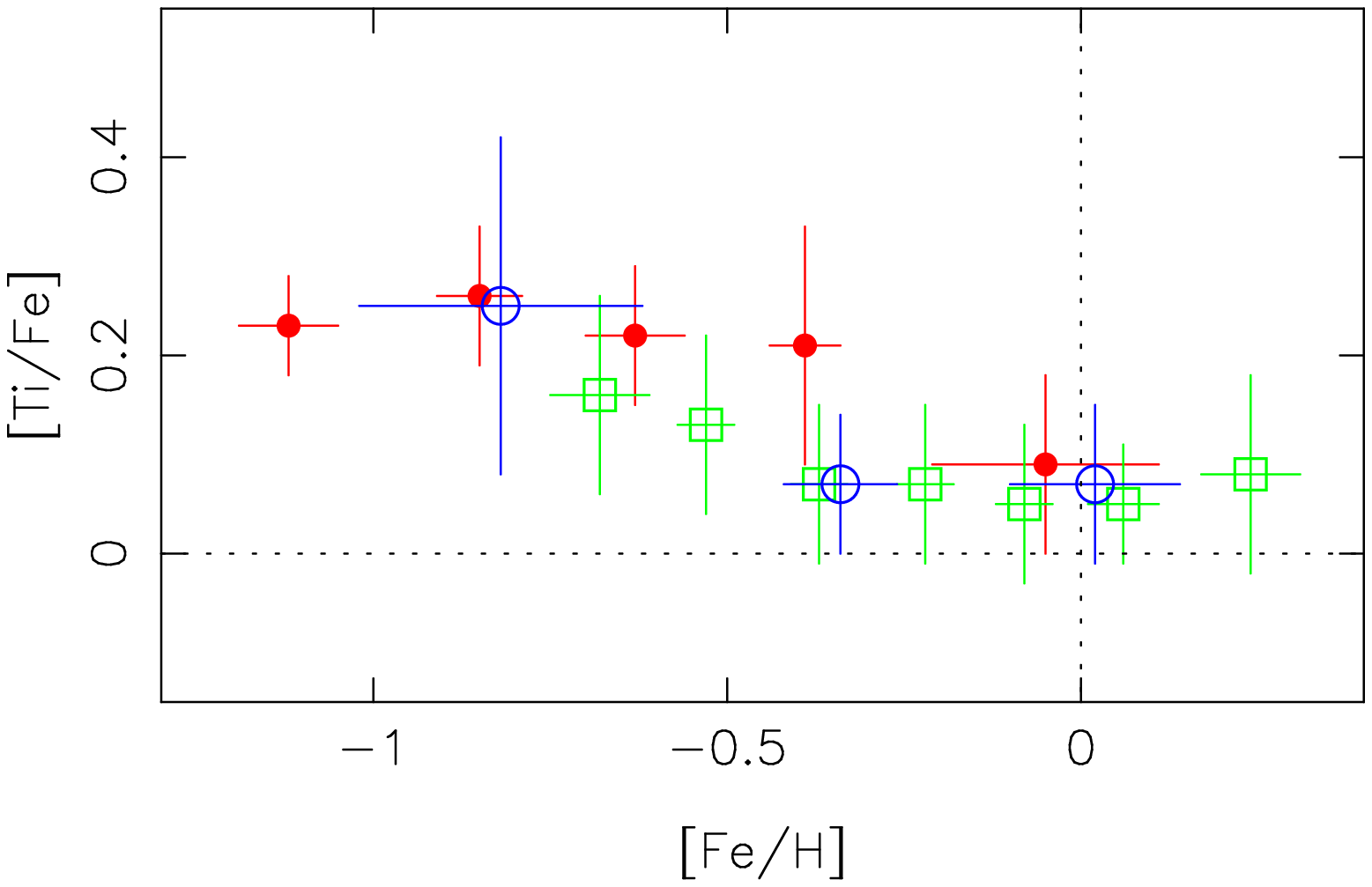}
\includegraphics[width=7cm]{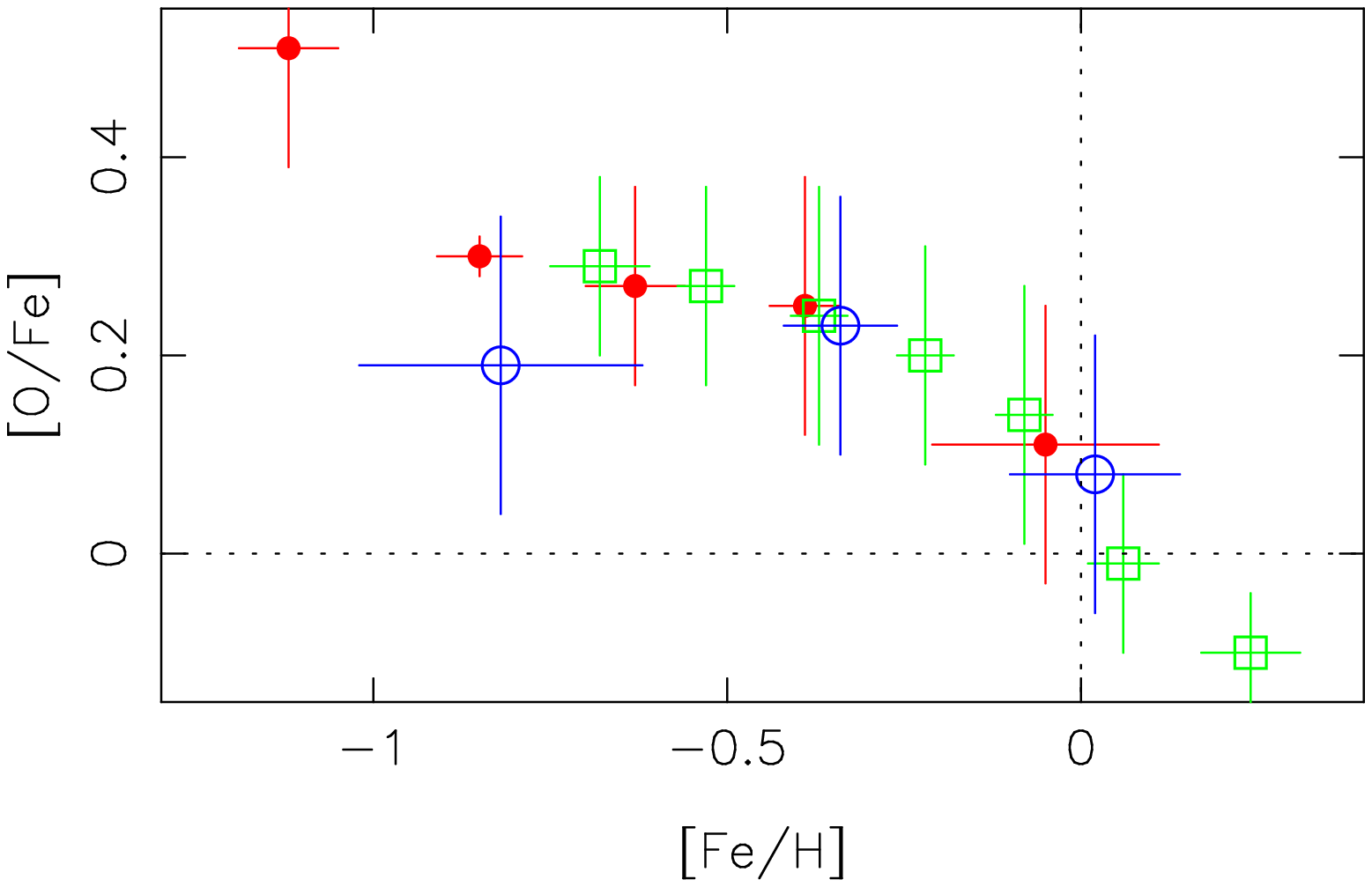}
\includegraphics[width=7cm]{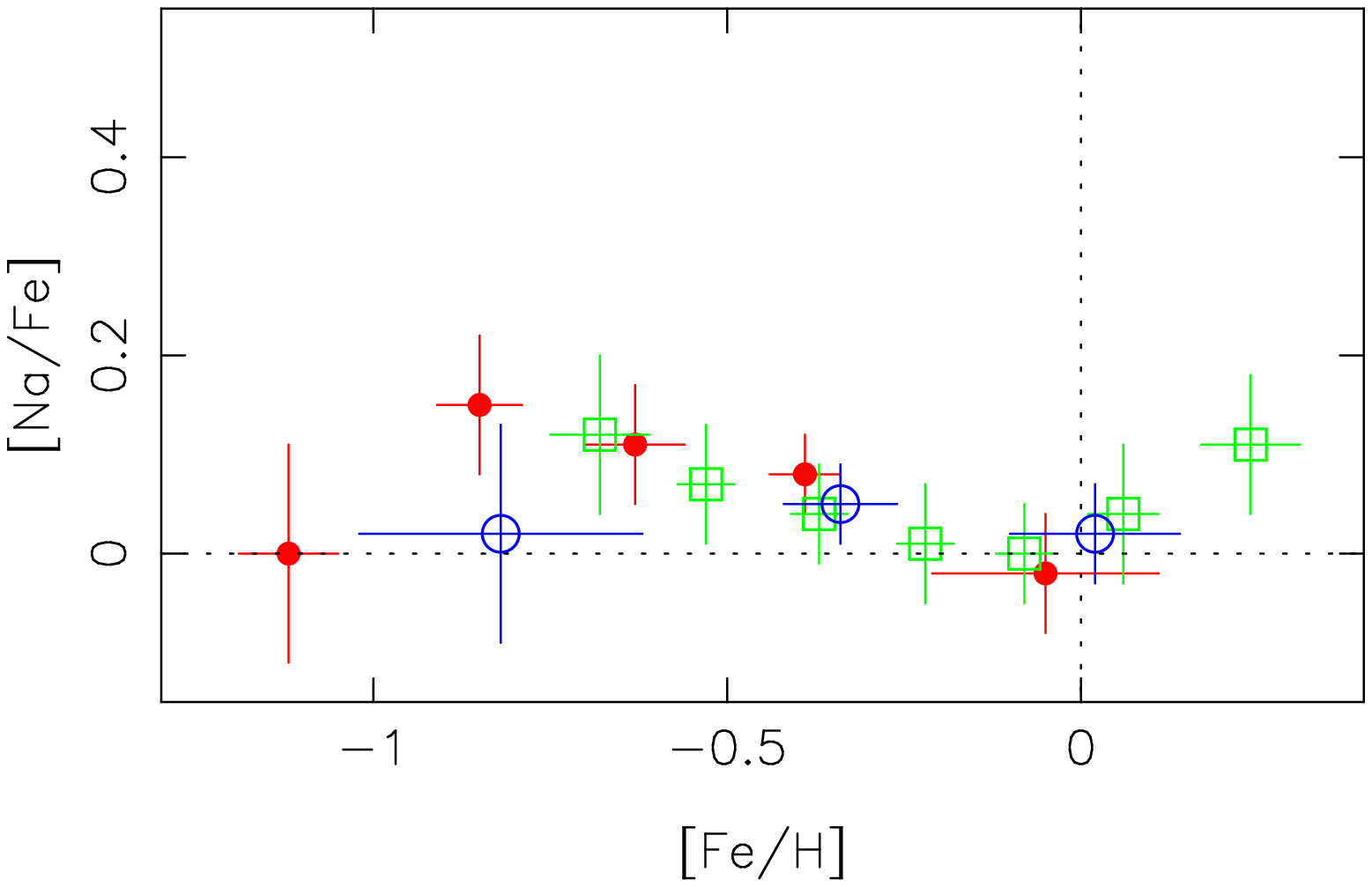}
\includegraphics[width=7cm]{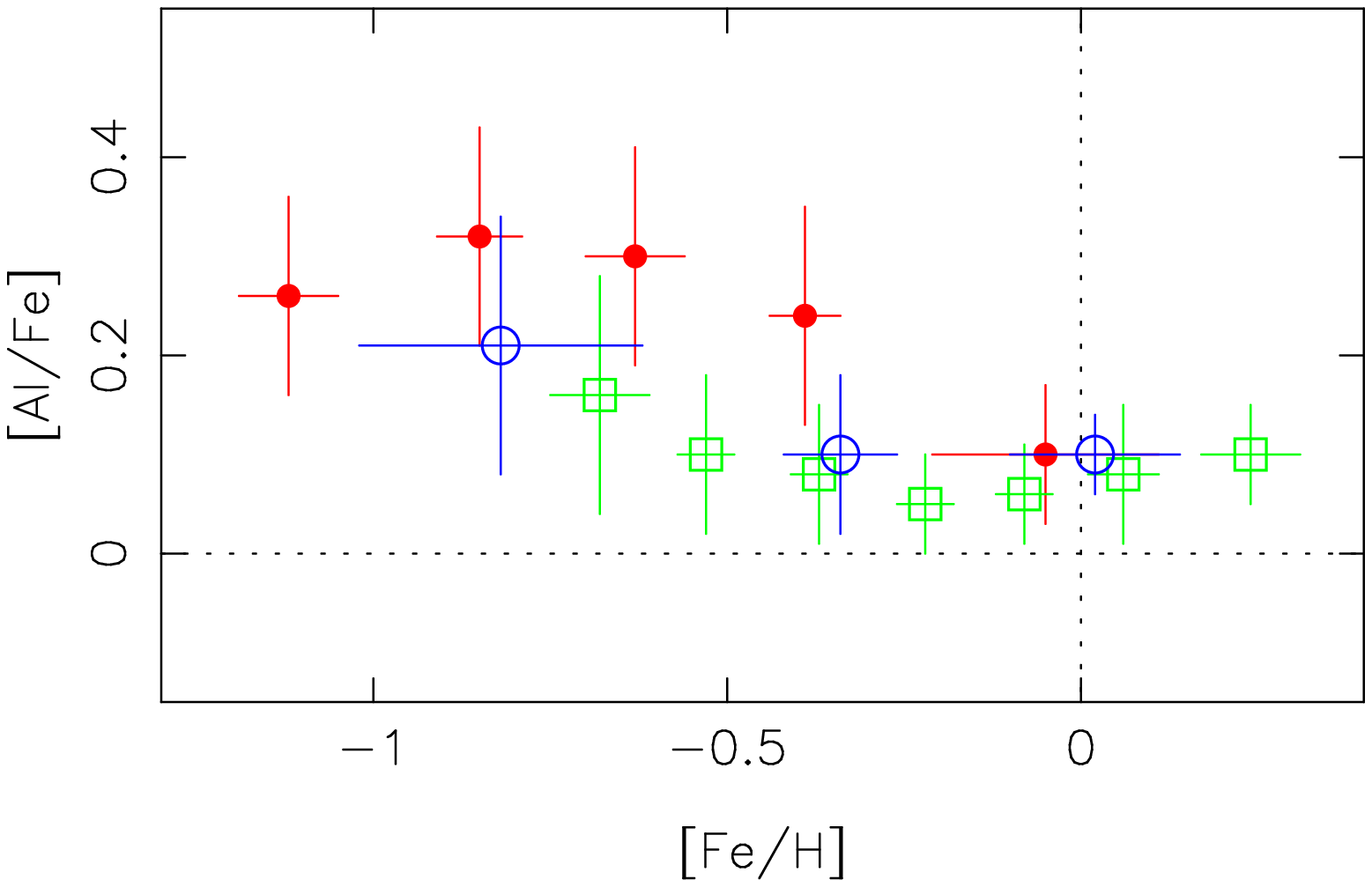}
\includegraphics[width=7cm]{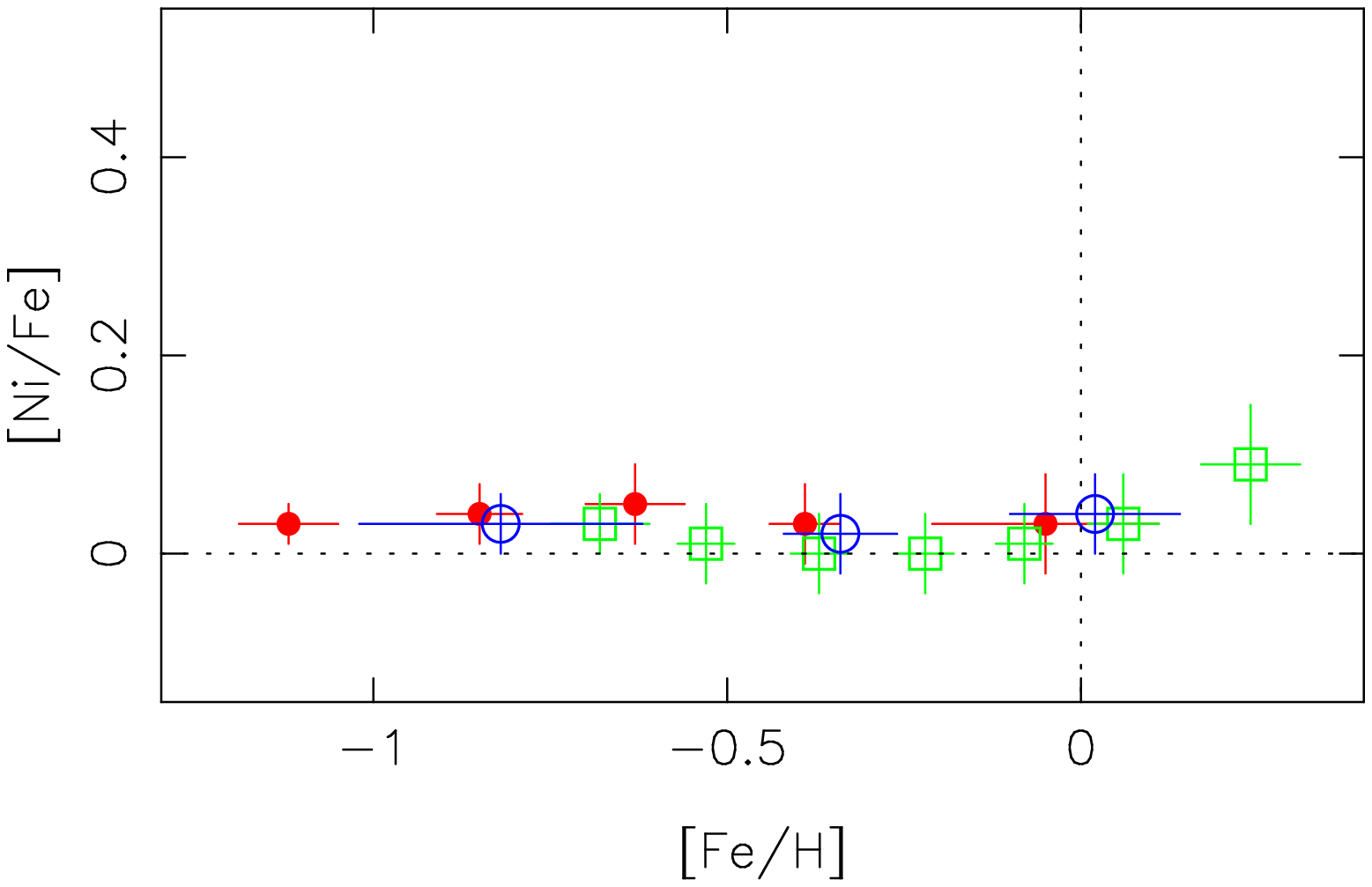}
\caption[]{Averaged [X/Fe] vs [Fe/H] per bin of metallicity in the thin disk (green squares)
in the thick disk (red filled circles) and in the Hercules stream (blue open circles).
Errors bars correspond to the standard deviations around the mean value in 
each bin.}
\label{f:bin_el_feh}
\end{figure*}

The abundance trends for each kinematical group are shown in 
Fig. \ref{f:bin_el_feh}. The number of stars involved in this
study has allowed us to bin the data in metallicity to highlight
the abundance features of each population.  Table \ref{t:bin_el} gives
the mean abundance, standard deviation and number 
of stars in each metallicity bin. The observed features are the following :
 
\begin{table*}[hbtp]
\caption{Mean abundance, standard deviation
and number of stars in each metallicity bin}
\label{t:bin_el}
\tiny
\begin{center}
\begin{tabular}{c r r r r r r r r r}
\hline
\hline
bin & [Fe/H] & [Si/Fe] & [Ca/Fe] & [Mg/Fe] & [Ti/Fe] & [O/Fe] &
[Na/Fe] & [Al/Fe] & [Ni/Fe] \\
\hline
\multicolumn{10}{c}{Thick disk}\\
$\lbrack$Fe/H$\rbrack <$ -1.00 & -1.12 &  0.35 &  0.29 &  0.45 &  0.23 &  0.51 &  0.00 &  0.26 &  0.03 \\
  &0.07 &  0.07 &  0.05 &  0.07 &  0.05 &  0.12 &  0.11 &  0.10 &  0.02 \\
  &   8 &     8 &     8 &     7 &     8 &     8 &     8 &     6 &     8 \\
 -1.00 $\le$ [Fe/H] $<$ -0.75 & -0.85 &  0.33 &  0.29 &  0.42 &  0.26 &  0.30 &  0.15 &  0.32 &  0.04 \\
  &  0.06 &  0.07 &  0.06 &  0.05 &  0.06 &  0.02 &  0.07 &  0.11 &  0.04 \\
  &    16 &    16 &    15 &    16 &    15 &     3 &    15 &    13 &    16 \\
 -0.75 $\le$ [Fe/H] $<$ -0.50 & -0.63 &  0.24 &  0.20 &  0.35 &  0.23 &  0.27 &  0.11 &  0.30 &  0.05 \\
  &  0.07 &  0.09 &  0.06 &  0.10 &  0.07 &  0.10 &  0.06 &  0.11 &  0.04 \\
  &    31 &    31 &    28 &    31 &    28 &    13 &    28 &    27 &    31 \\
 -0.50 $\le$ [Fe/H] $<$ -0.25 & -0.39 &  0.16 &  0.16 &  0.26 &  0.21 &  0.25 &  0.08 &  0.24 &  0.03 \\
  &  0.05 &  0.07 &  0.07 &  0.11 &  0.12 &  0.13 &  0.04 &  0.11 &  0.04 \\
  &    18 &    18 &    18 &    18 &    18 &    14 &    16 &    16 &    18 \\
 -0.25 $\le$ [Fe/H] & -0.05 &  0.07 &  0.03 &  0.09 &  0.09 &  0.11 & -0.02 &  0.10 &  0.03 \\
  &  0.16 &  0.05 &  0.06 &  0.09 &  0.09 &  0.14 &  0.06 &  0.07 &  0.05 \\
  &    11 &    11 &     9 &    11 &     9 &     8 &     8 &     8 &    11 \\
\hline
\multicolumn{10}{c}{Thin disk}\\
$\lbrack$Fe/H$\rbrack <$ -0.60 & -0.68 &  0.18 &  0.14 &  0.22 &  0.16 &  0.29 &  0.12 &  0.16 &  0.03 \\
  &  0.07 &  0.10 &  0.06 &  0.10 &  0.10 &  0.09 &  0.08 &  0.12 &  0.03 \\
  &    21 &    21 &    21 &    19 &    21 &    14 &    20 &    16 &    21 \\
 -0.60 $\le$ [Fe/H] $<$ -0.45 & -0.53 &  0.12 &  0.08 &  0.15 &  0.13 &  0.27 &  0.07 &  0.10 &  0.01 \\
  &  0.04 &  0.07 &  0.05 &  0.06 &  0.09 &  0.10 &  0.06 &  0.08 &  0.04 \\
  &    37 &    37 &    35 &    36 &    33 &    23 &    33 &    33 &    37 \\
 -0.45 $\le$ [Fe/H] $<$ -0.30 & -0.37 &  0.07 &  0.06 &  0.10 &  0.07 &  0.24 &  0.04 &  0.08 &  0.00 \\
  &  0.04 &  0.04 &  0.05 &  0.07 &  0.08 &  0.13 &  0.05 &  0.07 &  0.04 \\
  &    86 &    86 &    81 &    85 &    80 &    65 &    79 &    77 &    85 \\
 -0.30 $\le$ [Fe/H] $<$ -0.15 & -0.22 &  0.06 &  0.04 &  0.06 &  0.07 &  0.20 &  0.01 &  0.05 &  0.00 \\
  &  0.04 &  0.05 &  0.05 &  0.06 &  0.08 &  0.11 &  0.06 &  0.05 &  0.04 \\
  &    97 &    97 &    82 &    96 &    80 &    57 &    71 &    70 &    97 \\
 -0.15 $\le$ [Fe/H] $<$ 0 & -0.08 &  0.06 &  0.04 &  0.04 &  0.05 &  0.14 &  0.00 &  0.06 &  0.01 \\
  &  0.04 &  0.04 &  0.05 &  0.06 &  0.08 &  0.13 &  0.05 &  0.05 &  0.04 \\
  &    72 &    72 &    53 &    69 &    50 &    34 &    38 &    39 &    71 \\
 0 $\le$ [Fe/H] $<$ 0.15 &  0.06 &  0.06 &  0.03 &  0.06 &  0.05 & -0.01 &  0.04 &  0.08 &  0.03 \\
  &  0.05 &  0.04 &  0.05 &  0.06 &  0.06 &  0.09 &  0.07 &  0.07 &  0.05 \\
  &    83 &    83 &    65 &    81 &    65 &    38 &    53 &    49 &    82 \\
  0.15 $\le$ [Fe/H] & 0.24 &  0.10 &  0.04 &  0.06 &  0.08 & -0.10 &  0.11 &  0.10 &  0.09 \\
  &  0.07 &  0.06 &  0.07 &  0.10 &  0.10 &  0.06 &  0.07 &  0.05 &  0.06 \\
  &    32 &    32 &    27 &    32 &    27 &    14 &    19 &    19 &    32 \\
\hline
\multicolumn{10}{c}{Hercules stream}\\
$\lbrack$Fe/H$\rbrack <$ -0.50 & -0.82 &  0.23 &  0.20 &  0.32 &  0.25 &  0.19 &  0.02 &  0.21 &  0.03 \\
  &  0.20 &  0.09 &  0.09 &  0.08 &  0.17 &  0.15 &  0.11 &  0.13 &  0.03 \\
  &     8 &     8 &     8 &     8 &     8 &     4 &     6 &     3 &     8 \\
 -0.50 $\le$ [Fe/H] $<$ -0.20 & -0.34 &  0.11 &  0.06 &  0.14 &  0.07 &  0.23 &  0.05 &  0.10 &  0.02 \\
  &  0.08 &  0.07 &  0.05 &  0.09 &  0.07 &  0.13 &  0.04 &  0.08 &  0.04 \\
  &    17 &    17 &    13 &    15 &    13 &    11 &    12 &    11 &    17 \\
 -0.20 $\le$ [Fe/H] &  0.02 &  0.07 &  0.04 &  0.04 &  0.07 &  0.08 &  0.02 &  0.10 &  0.04 \\
  &  0.12 &  0.03 &  0.07 &  0.05 &  0.08 &  0.14 &  0.05 &  0.04 &  0.04 \\
  &    19 &    19 &    11 &    18 &    11 &     6 &     7 &     7 &    19 \\
\hline
\end{tabular}
\end{center}
\end{table*}

\begin{itemize}
\item the thin disk and the thick disk overlap greatly in metallicity : there are stars
with thin disk kinematics down to [Fe/H]=-0.80 and there are stars with 
thick disk kinematics at solar metallicity and above, with one of them in the 
super metal rich regime ([Fe/H] $>$ +0.25).
\item  the $\alpha$ elements (Mg, Si, Ti and Ca), as well as Al show the same behaviour :
at solar metallicity the three populations are confounded, from [Fe/H]$\simeq$ -0.3
to [Fe/H]$\simeq$ -0.7, the thin and thick disks show parallel trends with
an enhancement of these elements with respect to Fe in the thick disk
\item  in the range -0.70 $<$ [Fe/H] $<$ -0.30, [Mg/Fe] and [Al/Fe] of the thick disk
exceed that of the thin disk by $\sim$ 0.15 dex
\item a change of slope (a "knee") is visible in the thick disk 
for Si and Ca, less clearly for Mg, at [Fe/H] $\simeq$ -0.7
\item the only $\alpha$ element showing a knee in the thin disk is O at 
[Fe/H] $\simeq$ -0.5, the decrease being steep above this value
\item for O the three populations are confounded but the most metal-poor bin exhibits a 
large enhancement of [O/H]=+0.52
\item for Na and Ni the three populations are confounded
\item in the range -0.60 $<$ [Fe/H] $<$ -0.20, [Na/Fe] declines from $\sim$ +0.10 dex
to $\sim$ 0.00 dex
\item super metal rich stars of the thin disk exhibit a pronounced overabundance 
of Na and Ni
\item there are Hercules stars in the whole range of metallicity, which exhibit
abundances trends  similar to those of the thin disk
\end{itemize}

The overlapping metallicity distributions of the thin disk and the thick disk
has been previously established (B03, M04). A question which is not yet answered 
is where 
the metallicity distribution of two disks stop. Concerning the metal-poor side 
of the thin disk, we find 47 stars
with [Fe/H]  $<$ -0.50 among the 428 thin disk stars, the lowest value ([Fe/H]=-0.83)
being reached by HD134169.  Our sample is biased and may exagerate the fraction of 
metal-poor thin disk stars, but the existence of such stars 
is real and cannot be explained by measurement errors only. The super metal rich part
of the thin disk seems to have a different chemical behaviour than the rest of the 
thin disk, the most obvious difference being observed for Na and Ni. This raises 
the question wether these stars have the same origin as the other thin disk stars. 
Looking in detail at their velocity shows that half of the thin disk stars with
[Fe/H] $>$ +0.20 have a motion consistent with that of the Hyades-Pleiades 
supercluster. Famaey et al. (2004) propose this stream to have originated from 
a common large molecular cloud, radially perturbed by a spiral wave. Our observation of
similar high metallicities and peculiar abundance ratios for these stars is indeed
in good agreement with the hypothesis of their formation from common material.

Concerning the thick disk, our data clearly prove 
the existence of metal-rich high velocity stars. However there are arguments 
in favour and against the fact that the
thick disk extends at solar metallicity and even beyond. M04 have discussed this
issue and proposed the thick disk to stop at [Fe/H] $\simeq$ -0.30 on the basis
that more metal rich stars may belong to the Hercules stream (whose velocity
ellipsoid was not clearly defined at that time). Here we have carefully eliminated
Hercules stars from the thick disk sample but metal-rich stars remain. Moreover
Feltzing (2004) claims to have observed metal-rich stars at a large distance 
above the plane 
where the thick disk dominates. If the thick disk really extends at solar metallicity
then one has to explain why its $\alpha$  abundance trends are so different from
the thin disk, except at solar metallicity. The case of these metal-rich high velocity stars 
is further discussed in  the next section. 

The conclusion about the overlapping metallicity distributions 
of the thin and thick disks is that [Fe/H] is a very bad parameter by itself to
identify the disk populations. 

The $\alpha$ enhancement with respect to iron of the thick disk has been established a few years ago
by Gratton et
al. (1996) and Fuhrmann (1998) but evidence of the decreasing parallel trends 
for the thin and
thick disks is more recent and results from a careful kinematical deconvolution 
of the two disks by B03 and M04. The "knee", ie the
change of slope from a constant enhancement to a decreasing one, is interpreted by 
the typical signature of SNIa to the enrichment of the 
interstellar gas from which the later thick disk stars formed (Feltzing et al. 
2003). Our contribution
to this issue is to quantify the enhancement to be  +0.14 dex in Mg, a value
larger than the measurement errors. The enhancement of the other $\alpha$ elements,
 Si, Ca and Ti, is lower but their dispersion in the thick disk is also lower.
The very good correlation of Al and Mg has been discussed by E93 on the point of view of
yields of supernovae of different kind. On the point of view of galactic structure and
diagnosis to deconvolve the populations, [Al/Fe] turms to be a very good parameter. 
We have tested the efficiency of the parameters [Mg/Fe], 
[$\alpha$/Fe]=0.25([Mg/Fe]+[Si/Fe]+[Ca/Fe]+[Ti/Fe]) and [Mg+Al/Fe]=0.5([Mg/Fe]+[Al/Fe]) 
by performing linear regressions with [Fe/H]. We obtain the following relations 
(restricted to -0.80 $<$ [Fe/H] $<$ -0.30) :

For the thin disk :\\

[Mg/Fe]=-0.37[Fe/H]-0.040, $\sigma$=0.067 dex\\

[$\alpha$/Fe]=-0.29[Fe/H]-0.029, $\sigma$=0.052 dex\\

[Mg+Al/Fe]=-0.29[Fe/H]-0.021, $\sigma$=0.061 dex\\

For the thick disk :\\

[Mg/Fe]=-0.41[Fe/H]+0.097, $\sigma$=0.092 dex\\

[$\alpha$/Fe]=-0.30[Fe/H]+0.071, $\sigma$=0.069 dex\\

[Mg+Al/Fe]=-0.30[Fe/H]+0.128, $\sigma$=0.095 dex\\

Comparing for each abundance ratio the offset ($\Delta$) between the thin and thick disks with
the total dispersion ($\sigma_T$), the most favourable ratio is  
[Mg+Al/Fe] with $\Delta=1.31\sigma_T$ ($\Delta$=0.148 dex, $\sigma_T$=0.113 dex), whereas
$\Delta=1.20\sigma_T$ ($\Delta$=0.137 dex, $\sigma_T$=0.114 dex) is obtained for [Mg/Fe]
and
$\Delta=1.16\sigma_T$ ($\Delta$=0.100 dex, $\sigma_T$=0.086 dex) for [$\alpha$/Fe].

 If there is  a jump in the enhancement of oxygen
at [Fe/H]$<$-1.0, mainly due to the contribution of G03, our data do not show any 
enhancement of [O/Fe]
in the thick disk similar to the other $\alpha$ elements, a feature which was
observed by B04a. As mentioned previously, the
dispersions are large and may reflect the larger uncertainties which affect O abundance
determinations as compared to other elements. 
There is however a clear knee at [Fe/H]=-0.5, the decrease of [O/Fe] being steeper at larger
metallicites which 
suggests an abrupt change in the relative role of SN I and SN II.

 The fact that the Hercules stream occupies a large range of metallicity 
with abundance trends similar to those of the thin disk is in good agreement 
with the dynamical hypothesis
described by Famaey et al. (2004) since the perturbation of a rotating bar is
supposed to affect non contemporary but essentially thin disk stars.

\section{AMR, vertical gradient, metal-rich stars}
\label{s:grad}
\begin{figure}[hbtp]
\includegraphics[width=7cm]{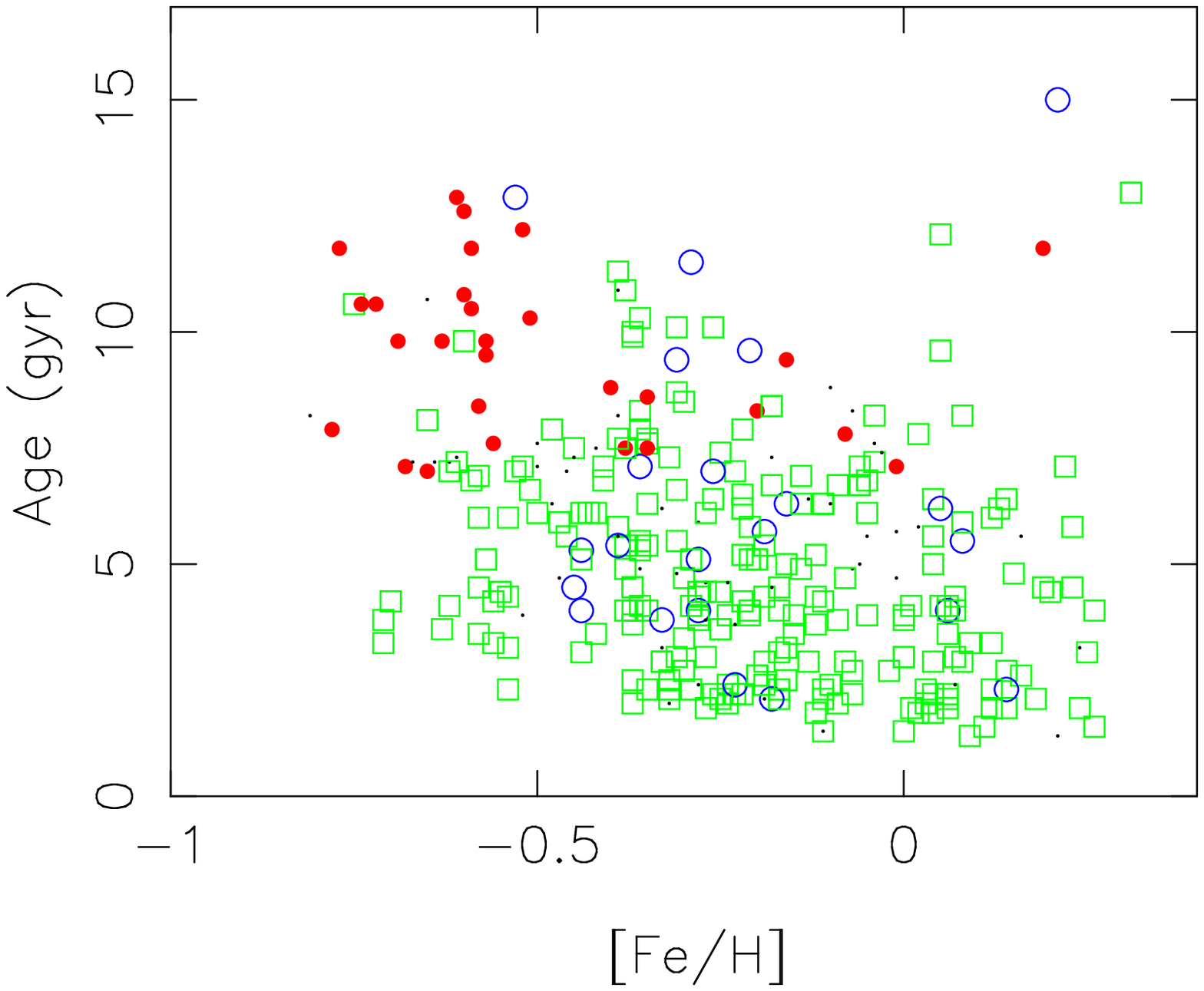}
\caption[]{Age vs [Fe/H] for stars having well defined ages. As in 
Fig. \ref{f:UV} and 
Fig. \ref{f:toom}, the different symbols indicate thin disk, thick disk and 
Hercules stars.}
\label{f:feh_age}
\end{figure}

\begin{figure}[hbtp]
\includegraphics[width=7cm]{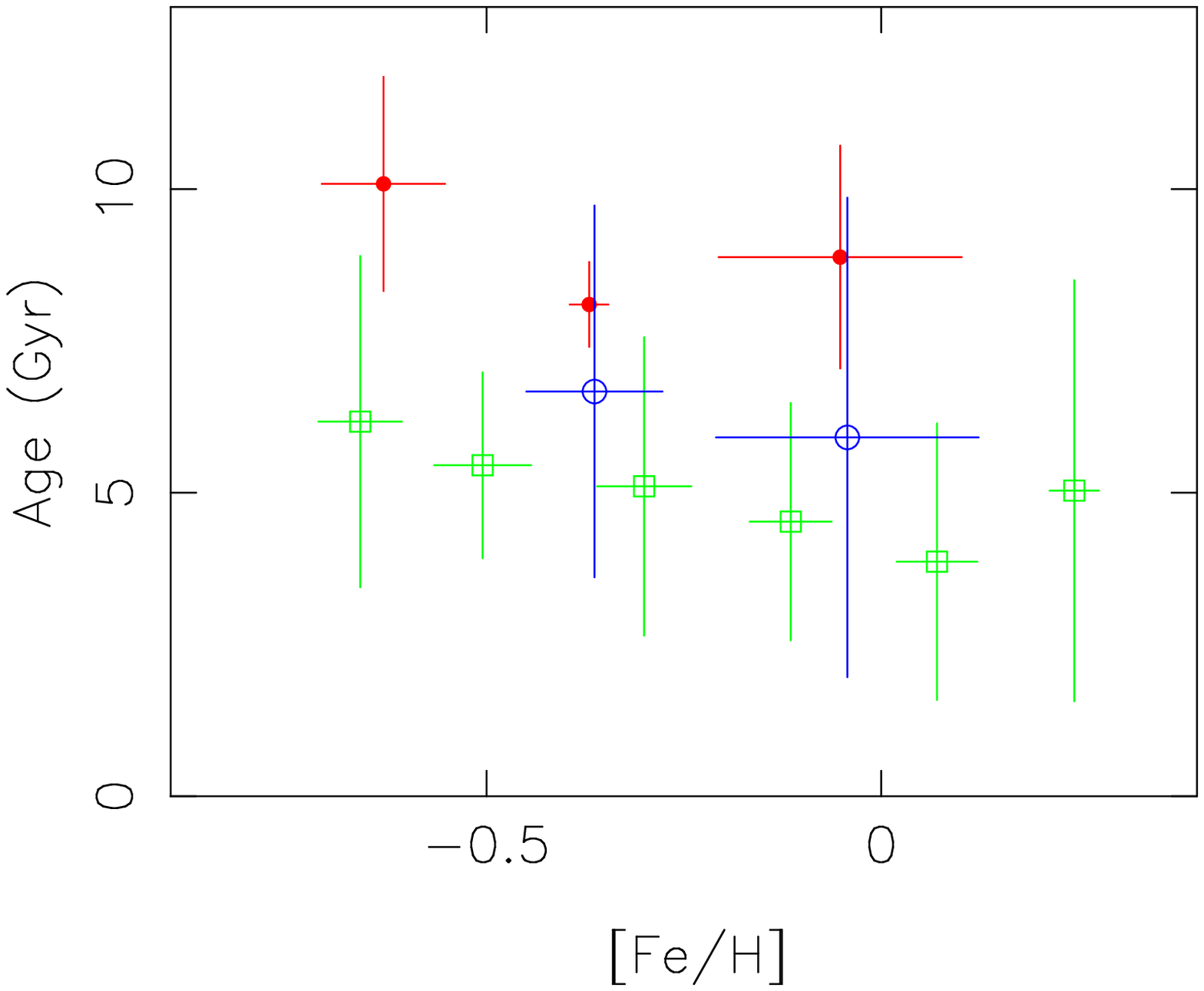}
\caption[]{Age vs [Fe/H] for stars having well defined ages. As in 
Fig. \ref{f:UV} and 
Fig. \ref{f:toom}, the different symbols indicate thin disk, thick disk and 
Hercules stars.}
\label{f:feh_age_bin}
\end{figure}

Fig. \ref{f:feh_age} and Fig. \ref{f:feh_age_bin} show that the thick disk 
is older than the thin disk in
the whole range of metallicity. Thick disk stars range from
7 to 13 Gyr with an average of 9.6 $\pm$ 0.3 Gyr (median 9.5 Gyr). Several thin disk stars
are found with ages greater than 10 Gyr, for instance HD127334 and HD190248 which
are also metal-rich and appear as outliers. An AMR is visible in the thin disk,
in the sense the most metal-poor thin disk stars are older than those with solar
metallicity by 2.3 Gyr on average.
The AMR in the thick disk is more difficult to establish : the statistic is poor, only
29 stars having an age determination. 
It depends also wether one classify the metal-rich stars into the
thick disk or not. Bensby et al. (2004b) have observed an AMR in the
thick disk from a larger sample kinematically selected, but with ages
determined from photometric metallicities. They obtain median ages of 13.0 Gyr at
[Fe/H]=-0.60 and 11.2 Gyr at [Fe/H]=-0.40. Our age scale is lower but 
we obtain a similar difference of age in the same metallicity bins (10.1 Gyr and
8.1 Gyr). The metal-rich bin of the thick disk does not follow the same trend. If one
consider this bin to be really part of the thick disk, then the AMR hypothesis is not
valid anymore. However we have several arguments against the metal rich high velocity
stars to be part of the thick disk (see below) and thus we interpret our data as
consistent with the existence of an AMR in the thick disk. Bensby et al. (2004b)
have estimated that star formation was active in the thick disk during 5 Gyr, but
this result is based on a wide metallicity distribution up to [Fe/H]=0. Our results
favour a shorter timescale for star formation, of the order of 2-3 Gyr.

Our data are also compatible with a hiatus in star formation between the formation 
of the thick disk and the thin disk. In the range -0.80 $<$ [Fe/H] $<$ -0.30, the 
mean age of the two disks differs by 4 Gyr. The reality of this offset is related to
the scatter of the age distribution in each population. The comparison 
with the Hercules stream is very interesting in that sense. According to Fig. \ref{f:feh_age},
the Hercules stream is made of stars of 
all ages. Consequently the age dispersion in its metallicity bins is very large. On the 
contrary, 
the thin and thick disks show much lower dispersions. The age dispersion in each bin 
results from the convolution of the cosmic 
scatter and the measurement errors of individual ages. We are aware that the uncertainty
on ages is large despite the
care we took to construct a clean sample. The median of the estimated errors is 2.7 Gyr.
This means that the cosmic scatter of ages in each metallicity bin is quite low
and the age offset between the two disks significant, pointing to a possible interruption
of star formation at the end of the thick disk formation and the beginning of the thin
disk formation. Incidentaly the large age dispersion of the Hercules stream favours the
dynamical hypothesis proposed by Famaey et al. (2004) as its origin.

\begin{figure}[hbtp]
\includegraphics[width=7cm]{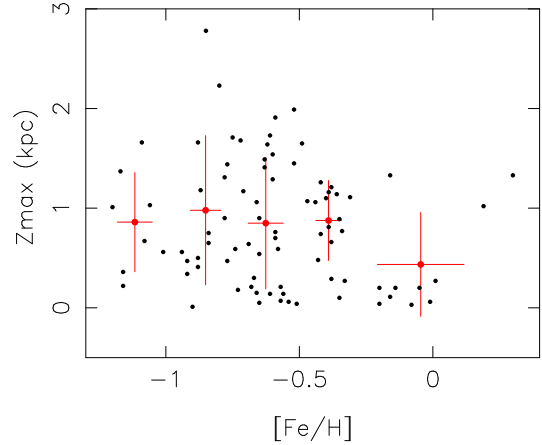}
\caption[]{Zmax, the maximal distance to the plane reached by the orbit, in several
bins of metallicity for thick disk stars. }
\label{f:fz}
\end{figure}

The existence of a vertical gradient of metallicity in the thick disk has important
consequence for the choice of the most probable scenario of its formation.
The relation between the maximal height above the plane, Zmax, and
the metallicity is shown in Fig. \ref{f:fz}. As previously mentioned in
Soubiran et al. (2003) and M04, a transition
occurs at [Fe/H]=-0.3. For stars more metal-poor than this value, the relation
Zmax vs [Fe/H] is flat, consistent with no vertical gradient. 

 The question wether the stars assigned to the 
thick disk in the highest metallicity bin 
 are real thick disk stars is important to clarify the extent of the 
metallicity distribution of the thick disk and the existence of an AMR.
It can be seen 
that thick disk stars with [Fe/H] $>$ -0.3 are of two
kinds : 8 stars have a flat distribution (Zmax $<$ 300 pc) which is
suspect for thick disk stars, whereas 3 stars appear as outliers with Zmax $>$
1 kpc. The 8 stars come from A04, B03, M04 and R03. Therefore their distribution
is not due to a peculiar selection bias of one of the eleven papers used to
construct our sample. The
3 outliers have also been studied by several authors : HD003628 by B03, F00 and
G03, HD145148 by B03 and HD190360 by B03 and M04. It is worth noticing that the 
combination of the eleven papers used to construct the catalogue of abundance did not fill the
gap of thick disk stars having [Fe/H] $>$ 0 and 300 pc $<$ Zmax $<$ 1 kpc.  Finally
stars with [Fe/H] $>$ -0.3 assigned to the thick disk differ from the rest of the thick 
disk on three points : their vertical distribution is inhomogeneous, their  
$\alpha$  abundances are similar to that of the thin disk, they do not follow the AMR
of the thick disk. Their nature has
still to be clarified with a complete sample. 

\section{Summary}
We have compiled a large catalogue of high quality stellar parameters of FGK stars 
to probe the properties
of the thin disk and the thick disk in the metallicity interval -1.30 $<$ [Fe/H] $<$ +0.50.
This catalogue includes 743 stars with abundance ratios of several $\alpha$ and 
iron peak elements, 639 stars with abundance ratios and accurate (U, V, W) velocity and orbits,
and 322 stars with abundance ratios, kinematical data and age estimations. All these
data are available in a form of a single electronic table at the CDS. We have used these data to
investigate the
metallicity, abundance ratios and age distributions of three kinematical groups with well defined
velocity ellipsoids. Three subsamples
have been selected on the basis of (U, V, W) velocities to be representative of
the thin disk (428 stars), of the thick disk (84 stars) and the Hercules stream (44 stars).

Our results confirm well established previous findings :
\begin{itemize}
\item The thin disk and the thick disk overlap in metallicity and exhibit parallel
slopes of [$\alpha$/Fe] vs [Fe/H] in the range -0.80 $<$ [Fe/H] $<$ -0.30, the thick disk being
enhanced.
\item The thick disk is older than the thick disk.
\end{itemize}

We bring new constraints on more controversial issues :
\begin{itemize}
\item The thin disk extents down to [Fe/H]=-0.80 and exhibits low dispersions in its abundance trends.
\item The thick disk also shows smooth abundance trends with low dispersions. The change
of slope
which reflects the contribution of the different supernovae to the ISM enrichment is visible
in [Si/Fe] vs [Fe/H] and [Ca/Fe] vs [Fe/H] at [Fe/H] $\simeq$ -0.70, less clearly in
[Mg/Fe] vs [Fe/H].
\item An AMR is visible in the thin disk, the most metal-poor stars having 6.2
Gyr on average, those with solar metallicity 3.9 Gyr.
\item  Ages in the thick disk range from
7 to 13 Gyr with an average of 9.6 $\pm$ 0.3 Gyr. There is a tentative evidence of an AMR
extending over 2-3 Gyr.
\item We do not find any evidence of a vertical metallicity gradient in the thick disk.
\item $\lbrack$O/Fe$\rbrack$ decreases in whole metallicity range with a change of
slope at [Fe/H]=-0.50.
\item  The most metal rich stars assigned
to the thin disk do not follow its global trends. They are significantly enhanced in all 
elements (particularly in Na and Ni) except
in O which is clearly depleted. They have also a larger dispersion in age. Half of these stars 
are probable members of the Hyades-Pleiades supercluster, two others are surprisingly 
old.
\end{itemize}

Finally we have also obtained new results :
\begin{itemize}
\item 
The slope of [$\alpha$/Fe] vs [Fe/H] in the interval -0.80 $<$ [Fe/H] $<$ -0.30 has been 
estimated for the two disks.
The enhancement
of $\alpha$ elements in the thick disk has been quantified to be +0.10 dex. [Mg/Fe] is more
efficient than [$\alpha$/Fe] to separate the two disks with an offset of +0.14 dex. Combining [Al/Fe] 
and [Mg/Fe] is
even better with an offset of +0.15 dex between the two disks and a lower dispersion.
\item The age difference between the thin  and  thick disks has been quantified to be 4 Gyr 
suggesting an interruption in star formation between their formation
\item The Hercules stream is found to span the whole metallicity and age range. Its 
chemical properties are similar to those of the thin disk.
This favours the dynamical hypothesis for its origin, related to the influence
of the central bar of the Galaxy which affects non contemporary stars.
\item Metal-rich stars assigned to the thick disk do not follow its global trends. The nature of 
these stars has to be clarified.
\end{itemize}

In this paper we have focussed on observational facts. The next step is to confront them 
with predictions of models. We believe that this dataset and our findings bring new
strong constraints to models of formation of the two components of the galactic disk,
and to models of chemical evolution.

\begin{acknowledgements}
We warmly thank Fr\'ed\'eric Pont who kindly computed the ages with his Bayesian method.
This  research  has made  use  of  the  SIMBAD and  VIZIER  databases,
operated at CDS, Strasbourg,  France. It has used data from  the ESA 
{\it  Hipparcos} satellite (Hipparcos and Tycho--2 catalogues).
\end{acknowledgements}

\end{document}